\documentstyle[11pt]{article}
\def\be{\begin{equation}}

\def\ee{\end{equation}}

\def\etal{{\em et al.}}
\def\hkpc{h^{-1}{\rm kpc}}
\def\hMpc{h^{-1}{\rm Mpc}}
\def\h3Mpc{h^{-3}{\rm Mpc}^3}

\def\h3Mpcinv{h^{3}{\rm Mpc}^{-3}}
\def\ie{{\em i.e.}\ }
\def\kms{\mbox{km s$^{-1}$}}

\def\spose#1{\hbox to 0pt{#1\hss}}
\def\simlt{\mathrel{\spose{\lower 3pt\hbox{$\mathchar"218$}}
     \raise 2.0pt\hbox{$\mathchar"13C$}}}
\def\simgt{\mathrel{\spose{\lower 3pt\hbox{$\mathchar"218$}}
     \raise 2.0pt\hbox{$\mathchar"13E$}}}

\def\inserthang{\everypar={\parindent=0pt\hangindent=1.5pc\hangafter=1}}
\def\refmode{\parindent=0pt\inserthang\parskip 0pt
\frenchspacing\exhyphenpenalty=10000\hyphenpenalty=10000}
\def\refer#1{#1}
\def\refeq#1{\relax}		
\def\IAU130{in IAU Symp. 130, Large Scale Structures of the Universe}

\input epsf
\begin{document}

\begin{center}
 {\large\bf The Stromlo-APM Redshift Survey III.\\
            Redshift Space Distortions, Omega and Bias}
\\[0.5in]
	{\bf J.~Loveday}\\
	Fermi National Accelerator Laboratory, PO Box 500, Batavia, 
	IL 60510, USA\\
	loveday@fnal.gov\\[0.1in]
	{\bf G.~Efstathiou}\\
	Department of Physics, Keble Road,
	Oxford, OX1 3RH, England\\
	g.efstathiou@physics.oxford.ac.uk\\[0.1in]
	{\bf S.J.~Maddox}\\
	Royal Greenwich Observatory, Madingley Road, Cambridge, 
	CB3 0EZ, England\\
	sjm@mail.ast.cam.ac.uk\\[0.1in]
	{\bf B.A.~Peterson}\\
	Mount Stromlo and Siding Spring Observatories,
	Weston Creek PO, ACT 2611, Australia\\
	peterson@mso.anu.edu.au\\[0.2in]

 Revised February 26, 1996
\end{center}
\section*{Abstract}

Galaxy redshift surveys provide a distorted picture of the universe
due to the non-Hubble component of galaxy motions.
By measuring such distortions in the linear regime
one can constrain the quantity
$\beta = \Omega^{0.6}/b$ where $\Omega$ is the cosmological density
parameter and $b$ is the (linear) bias factor for optically-selected galaxies.
In this paper we apply two techniques for estimating $\beta$ from the
Stromlo-APM redshift survey --- (1) measuring the anisotropy of
the redshift space correlation function in spherical harmonics and 
(2) comparing the amplitude
of the direction-averaged redshift space correlation function to the
real space correlation function.
We test the validity of these techniques, particularly whether the assumption
of linear theory is justified, using two sets of large $N$-body simulations.
We find that the first technique is affected by non-linearities on scales
up to $\sim 30 \hMpc$.
The second technique is less sensitive to non-linear effects and so is
more useful for existing redshift surveys.

The Stromlo-APM survey data favours a low value for $\beta$, with
$\beta \simlt 0.6$.
A bias parameter $b \approx 2$ is thus required if $\Omega \equiv 1$.
However, higher-order correlations measured from the APM galaxy survey
(Gazta\~{n}aga and Frieman 1994) indicate a low value for the bias
parameter $b \approx 1$, requiring that $\Omega \simlt 0.5$.
We also measure the relative bias for samples of galaxies of various luminosity
and morphological type and find that low-luminosity galaxies are roughly
three times less biased than $L^*$ galaxies.
For the galaxy population as a whole, we measure a real space
variance of galaxy counts
in $8 \hMpc$ spheres of $(\sigma^2_8)_g = 0.89 \pm 0.05$.

Subject headings: galaxies: clustering --- galaxies: distances and redshifts
--- galaxies: fundamental parameters
--- large-scale structure of universe --- surveys

\newpage
\section{Introduction}

Galaxy redshift surveys can provide some of the most important constraints on 
theories of large-scale structure, but they must be analysed with care.
For pure, unperturbed Hubble flow, galaxy clustering measured in 
redshift space would be isotropic and identical to that measured in
real space.
However, in practice, peculiar velocities will distort the redshift
space correlation function.
Since the amplitude of peculiar velocities depends on the cosmological
density parameter $\Omega$, measurements of
this distortion can constrain the value of $\Omega$.
On small scales, the effect of peculiar velocities is to elongate
clusters of galaxies along the line of sight in redshift space,
leading to the well known `fingers of God'.
However, on large scales, coherent bulk flows dominate the peculiar
velocity field resulting in a {\em compression} in the clustering pattern
along the line of sight.
This effect is easily seen in the Stromlo-APM Survey as shown
in Figure~1, which is a contour plot of
the full redshift space correlation function
$\xi(\sigma,\pi)$ 
as a function of components of separation parallel ($\pi$) and
perpendicular ($\sigma$) to the line of sight.
A compression of the low-amplitude $\xi$ contours in the $\pi$ direction
compared with the $\sigma$ direction is clearly visible for
$\pi \simgt 10 \hMpc$.
(We assume a Hubble constant of $H_0 = 100 h$ km/s/Mpc.)

This large-scale anisotropy in redshift space clustering is most naturally
expressed in terms of the power spectrum.
\refer{Kaiser (1987)} has shown that in the linear regime of gravitational
instability models, the power spectra in redshift space, $P_s(\bf k)$
and real space $P_r(\bf k)$ are simply related by
\be
  P_s({\bf k}) = (1 + \beta \mu_{\bf k}^2)^2 P_r({\bf k}),
\ee
where $\mu_{\bf k}$ is the cosine of the angle between the wavevector $\bf k$
and the line of sight.
The amplitude of the distortion is determined by the parameter
$\beta = f(\Omega)/b$,
where $f(\Omega) \approx \Omega^{0.6}$ is the dimensionless 
growth rate of growing modes in linear theory.
The bias parameter $b$ relates the fluctuations in galaxy density
to the underlying mass density in the linear regime,
$\delta_g = b \delta_\rho$ for linear bias.
Several practical methods for measuring $\beta$ have recently been
applied: measuring the anisotropy of the correlation function
(\refer{Hamilton 1992}, 1993a\refeq{Hamilton 1993a}; 
\refer{Fisher \etal\ 1994a}), the anisotropy of
the power spectrum (\refer{Cole, Fisher and Weinberg 1994}, 
1995\refeq{Cole, Fisher and Weinberg 1995};
\refer{Tadros and Efstathiou 1996})
and spherical harmonics of the density field 
(\refer{Fisher \etal\ 1994b};
\refer{Heavens and Taylor 1995}).

We follow the correlation function approach in this paper.\footnote{
For a power spectrum analysis of the Stromlo-APM survey, see 
Tadros and Efstathiou (1996).}
\refer{Hamilton (1992)} has pointed out that the cosine $\mu_{\bf k}$
in Fourier space transforms to an operator in real space:
\be
 \xi_s(r,\mu) = [1 + \beta (\partial/\partial z)^2 (\nabla^2)^{-1}]^2 \xi_r(r),
\ee
where $(\nabla^2)^{-1}$ denotes the inverse Laplacian operator
and $\mu = \hat{\bf r}.\hat{\bf z}$
is the cosine angle between pair separation $\bf r$ and the line of sight 
$\bf z$.
The redshift space correlation function $\xi_s(r,\mu)$
is conveniently expressed as a sum of spherical harmonics involving
the first three even-order Legendre polynomials,
(the odd-order harmonics vanish by pair-exchange symmetry)
\be
  \xi_s(r,\mu) = \xi_0(r) P_0(\mu) + \xi_2(r) P_2(\mu) + \xi_4(r) P_4(\mu).
\ee
Hamilton gives expressions for the $\xi_l(r)$ in terms of integrals over
$\xi_r(r)$ [his equations (6)--(9)].

To solve the inverse problem, \ie to go from redshift space clustering
to real space clustering, Hamilton integrates the equations describing
the real and redshift space correlation functions over planes
normal to the vector $\bf r$ at separation $r$, expands in spherical
harmonics and differentiates with respect to $r$.
He thereby obtains an
explicit expression for $\beta$ and $\xi_r(r)$ in terms of the harmonics $\xi_l(r)$
of the redshift space correlation function.
Further volume averaging to minimize cancelation of terms finally results
in an equation for $\beta$ involving the 0th and  2nd order harmonics
of the redshift space correlation function:
\be
  \frac{1 + \frac{2}{3}\beta + \frac{1}{5}\beta^2}
       {\frac{4}{3}\beta + \frac{4}{7}\beta^2}
  = \frac{\xi_0(r) - 3 \int_0^r \xi_0(s) (s/r)^3 ds/s}{\xi_2(r)}.
  \label{eqn:beta_02}
\ee
One can also write a similar expression involving the 2nd and 4th order
harmonics, but in practice, the 4th order harmonic
is too noisy and too strongly affected by non-linear effects to be useful.

As well as causing anisotropy in redshift space, large-scale streaming
motions also produce an amplification in the direction-averaged
redshift space correlation function on large scales.
For fluctuations in the linear regime, the direction-averaged
redshift space correlation
function $\xi(s)$ and the real space correlation function $\xi(r)$
are related by (Kaiser 1987)
\be
\xi(s) \approx \left( 1 + \frac{2}{3} \beta + \frac{1}{5}\beta^2 \right)\xi(r).
\label{eqn:beta_sr}
\ee

The large uncertainty in the value of $\beta$ hinders comparison
of $\xi(s)$ with real space predictions of galaxy clustering
from various models (see, for example, \refer{Loveday \etal\ 1992a}).
In a recent paper (\refer{Loveday \etal\ 1995}, hereafter Paper~2), 
we estimated the real-space correlation function
of optically-selected galaxies by cross-correlating galaxies in the
sparse-sampled Stromlo-APM Redshift Survey with the fully-sampled, parent
APM Galaxy Survey.
This projected cross-correlation function is unaffected by redshift-space
distortions and may be stably inverted to give the real-space correlation
function $\xi(r)$.
Moreover, the large number of cross-pairs enables clustering to be measured
to smaller scales than using the redshift survey data alone.
If both $\xi(r)$ and $\xi(s)$ can be reliably measured in the linear regime
then the value of $\beta$ can be constrained using 
equation~(\ref{eqn:beta_sr}).

We assess the relative merits of these two methods of estimating
$\beta$ (equations \ref{eqn:beta_02} and \ref{eqn:beta_sr})
by analyzing $N$-body simulations.  The two sets of simulations,
of low and high density, are designed
to be similar to the Stromlo-APM data in both the clustering and
dynamics of the galaxies.  In particular, we estimate the sensitivity
of the methods to non-linear dynamics of the galaxy distribution.

The above expressions (1--5) assume a plane-parallel approximation for 
peculiar displacements.
In order to approximate this ideal in our analyses,
we use only those pairs of galaxies
separated by less than 50 degrees on the sky.
This rejects about 20\% of galaxy pairs, and, as Cole \etal\ (1994) 
have demonstrated, will limit deviations from the plane-parallel
approximation to no more than a 5\% bias in the estimated value of $\beta$.

Throughout the paper, we use $r$ to denote real space separations
and $s$ to denote separations in redshift space.
Error bars on measurements from survey data are estimated using the
bootstrap resampling technique (\refer{Barrow, Bhavsar and Sonoda 1984})
with nine bootstrap resamplings of the survey.
Error bars for simulations are determined from the variance between
ten independent realizations of the low-density model and nine
realizations of the high-density model.

The layout of the paper is as follows.
The Stromlo-APM survey data and the $N$-body simulations 
are described and compared in \S\ref{sec:survey}.
In \S\ref{sec:tests} we test the estimators for $\beta$ using the simulations.
In \S\ref{sec:results} we apply the estimators to the Stromlo-APM data and
also present
the relative bias and redshift space distortions for different galaxy types.
Finally, our conclusions are given in \S\ref{sec:concs}.

\section{Survey Data and N-Body Simulations}
\label{sec:survey}

Both equations~(\ref{eqn:beta_02}) and (\ref{eqn:beta_sr})
assume that linear theory is valid on scales on which $\xi$ can be
reliably measured.
It is important to test this assumption of linearity before using these
equations to estimate $\beta$.
We do this by analyzing two ensembles of CDM-like $N$-body simulations.
In this section we describe the Stromlo-APM survey data and the simulations,
and show that the simulations mimic the observed clustering and dynamics
of the galaxy distribution quite faithfully.

\subsection{Stromlo-APM Survey Data}

The Stromlo-APM redshift survey consists of 1787 galaxies with $b_J \le
17.15$ selected randomly at a rate of 1 in 20 from the APM (Automated
Plate Measuring) galaxy survey
(Maddox \etal\ 1990a,b).
\refeq{Maddox etal 1990a}\refeq{Maddox etal 1990b}
The survey covers a solid angle of 1.3 sr (4300 square degrees)
in the south galactic cap.
The APM magnitudes have been calibrated and corrected for photographic
saturation using CCD photometry as described by \refer{Loveday \etal\ 1992b}
(hereafter Paper~1).
An approximate morphological type was assigned to each galaxy by visually
inspecting the images on the United Kingdom Schmidt Telescope (UKST) 
survey plates. 
Redshifts were obtained with the Mount Stromlo-Siding Spring Observatory
(MSSSO) 2.3m telescope at Siding Spring.  
Measured radial velocities were transformed to the local group frame using
$v = v + 300\sin(l)\cos(b)$ and we assumed $\Lambda = 0$, $q_0 = 0.5$ and
$H_0 = 100$ \kms Mpc$^{-1}$ with uniform Hubble flow in calculating
distances and absolute magnitudes.  
We adopt $k$-corrections for different morphological types in the $b_J$
system as described by \refer{Efstathiou, Ellis and Peterson (1988)}. 
More details about the survey are given in Paper~1 and
the construction of the survey
is described in full by \refer{Loveday \etal\ (1996)}.

\subsection{Simulations}
\label{sec:sims}

We use the two sets of $N$-body simulations described by
\refer{Croft and Efstathiou (1994)}; a low-density CDM model with
$\Omega_0 = 0.2$ and a cosmological constant (LCDM) and a mixed
dark matter model with $\Omega_0 = 1$ (MDM).
We analyze ten realizations of the LCDM model and nine realizations of the 
MDM model.
These simulations combine a large volume (box length = $300 \hMpc$)
with a force resolution of $\approx 80 \hkpc$ for $10^6$ particles, and so can
be used to generate reasonable approximations to our redshift survey.
Both sets of $N$-body simulations have enhanced large-scale power compared
with the standard CDM model; the LCDM model by having 
$\Gamma = \Omega_0 h = 0.2$ and a nonzero
cosmological constant $\lambda = \Lambda/(3 H_0^2) = (1 - \Omega_0) = 0.8$
and the MDM model by having $\Omega_\nu = 0.3$ from neutrinos and $h = 0.5$.

In order to generate mock Stromlo-APM catalogues from the simulations,
`galaxies' were selected within the APM area and with the Stromlo-APM 
selection function (Paper~1), such that they traced the mass
particles in an unbiased way.
This procedure produces mock catalogues of on average 33,500 `galaxies' each.
We select a subset of 1 in 20 galaxies at random from each simulated catalogue
in order to mimic the sparse sampling strategy of the Stromlo-APM survey.
The intrinsic value of $\beta$ for the simulations is, by construction,
$\beta = 0.38$ for the LCDM simulations and $\beta = 1$ for the MDM 
simulations.

\subsection{Galaxy Clustering in Real and Redshift Space}

We estimate the redshift-space correlation functions from the survey and the
simulations using the density-independent estimator for $\xi(s)$ discussed in
Paper~2,
\be
  1 + \xi(s) = \frac{w_{gg}(s) w_{rr}(s)}{[w_{gr}(s)]^2}.
  \label{eqn:xi_ni}
\ee
Here $w_{gg}(s)$, $w_{gr}(s)$ and $w_{rr}(s)$ are the summed products of 
weights of galaxy-galaxy, galaxy-random and random-random pairs respectively.
Note that the relative densities of galaxy and random points measured
at separation $s$ are automatically accounted for by this estimator ---
there is no need to assume an overall galaxy density $n_g$.
This estimator, due to \refer{Hamilton (1993b)}, is insensitive to variations
in galaxy density and provides a very stable estimate of $\xi(s)$.

To calculate the real-space correlation function $\xi(r)$,
we measure the projected cross-correlation function between the redshift
data and the angular data from the fully-sampled parent catalogue,
\be 
\Xi(\sigma) = \int_{-\infty}^{+\infty} 
\xi( \sqrt{ \Delta y^2 + \sigma ^2 } ) d \Delta y,
\label{eqn:xi_proj}
\ee
where the integral extends over all line-of-sight separations $\Delta y $ for
pairs of galaxies with given projected separation $ \sigma = y
\theta $ ($\theta$ is the angular separation and $y$ is the distance
to the galaxy of known redshift). 
This projected function is inverted numerically to
give an estimate of $\xi(r)$, which is unaffected by
redshift-space distortions.
See \refer{Saunders \etal\ (1992)} and Paper~2
for a detailed description of this estimator.

In Figure~\ref{fig:xi_apm_nsim} we compare the clustering of galaxies
in (a) redshift space and (b) real space for the Stromlo-APM survey and 
for the $N$-body simulations.
The error bars show the scatter
between nine bootstrap resamplings of the survey.
The measured variance between the different realizations of the simulations
yields similar error bars, which are not plotted here for the sake of clarity.
We see that the simulations match the observed clustering of
galaxies reasonably well on large scales, although on small scales
the LCDM models overpredict and the MDM models underpredict the observed
clustering.
The plateau in the real-space clustering of Stromlo-APM
galaxies at $r \approx 20$--$30 \hMpc$ appears to be significant
given the size of the random errors.
However, as discussed in Paper~2, 
$\xi(r)$ inferred from inversion of the projected
cross-correlation with the parent catalogue is subject to systematic error
beyond $r \approx 20 \hMpc$, although neither of the simulations shows
any feature here.
It is clear that we will need to check that we obtain consistent
estimates of $\beta$ on different scales if our results are to be completely
trustworthy.

We have fit power-laws to the correlation functions plotted in
Figure~\ref{fig:xi_apm_nsim}.
We fit over the range $1.5 < s < 30\ \hMpc$ in redshift space and
$0.2 < r < 20\ \hMpc$ in real space.
The parameters to these power-law fits are given in Table~\ref{tab:xi_fits}.

\subsection{Galaxy Peculiar Velocities}
\newcommand{\wrms}{\mbox{$\langle w^2 \rangle^{1/2}$}}

If the $N$-body simulations are to be used to check for non-linear effects
in the real data, then it is important to compare the amplitude of
small-scale, virialized motions in the simulations with those in the data.
By comparing estimates of clustering in redshift space and real space,
one can constrain the galaxy peculiar velocity distribution
$f(w)$ (eg. \refer{Bean \etal\ 1983}, \refer{Davis and Peebles 1983}).
The redshift space correlation function $\xi(\sigma,\pi)$ is given by
convolving the real space correlation function $\xi(r)$ with $f(w)$,
\be
  1 + \xi(\sigma,\pi) = \int_{-\infty}^{\infty} [1 + \xi(r)]
			f[w_3 + H_0 \beta \xi(1 + \xi)^{-1} r_3] dw_3,
  \label{eqn:xiv}
\ee
(Bean \etal\ 1983), where $r^2 = \sigma^2 + r_3^2$, $r_3 = \pi - w_3/H_0$
(the subscript 3 denotes the line-of-sight component of a vector quantity)
and $\langle w \rangle \approx -H_0 \beta \xi (1 + \xi)^{-1} r_3$ is the 
mean streaming velocity of galaxies at separation $r$.

We have measured $\xi(\sigma,\pi)$ in four $\sigma$ bins each of width
$2\hMpc$ centred on 1, 3, 5 and $7 \hMpc$ --- Figures~\ref{fig:v_pec}$a$, $b$
and $c$ for the data, LCDM and MDM simulations respectively.
We have calculated the best-fit rms peculiar velocity, \wrms, for three
models for $f(w)$, a Gaussian,
\be
  f(w) = \frac{1}{\sqrt{2\pi}\wrms} \exp\left(\frac{-w^2}{2\langle w^2 \rangle}
	 \right),
\ee
a $|w|^{3/2}$ distribution,
\be
  f(w) = \frac{0.476}{\wrms} \exp\left(\frac{-0.7966 |w|^{3/2}}
					    {\langle w^2 \rangle^{3/4}}\right),
\ee
and an exponential distribution,
\be
  f(w) = \frac{1}{\sqrt{2}\wrms} \exp\left(\frac{-\sqrt{2}|w|}{\wrms}\right).
\ee

The optimum value of \wrms\ for each distribution was calculated by
maximizing the likelihood
\be
  {\cal L} = \prod_{\pi-{\rm bins}} [2\pi {\rm Var}\{\xi(\sigma,\pi)\}]^{-1/2}
	\exp\left[\frac{-(\xi^o(\sigma,\pi) - \xi^p(\sigma,\pi))^2}
		       {2 {\rm Var}\{\xi(\sigma,\pi)\}} \right],
\ee
where $\xi^o(\sigma,\pi)$ is the observed redshift space correlation function,
${\rm Var}\{\xi(\sigma,\pi)\}$ is the observed variance in 
$\xi^o(\sigma,\pi)$
from bootstrap resampling (or from different realizations of the simulations)
and $\xi^p(\sigma,\pi)$ is the predicted
correlation function from (\ref{eqn:xiv}).
We use the measured $\xi(r)$ from Figure~\ref{fig:xi_apm_nsim}(b)
in equation (\ref{eqn:xiv}) and substitute $\beta = 0.38$
and $\beta = 1$ for the LCDM and MDM simulations respectively
and assume $\beta = 0.5$ for the Stromlo-APM data (see \S\ref{sec:results}).
The continuous, dashed and dot-dashed lines in Figure~\ref{fig:v_pec}
are the best fit curves for the Gaussian, $|w|^{3/2}$ and exponential
models respectively.
The best-fit values of \wrms\ together with
95\% confidence limits as estimated from likelihood ratios
are given in Table~\ref{tab:v_pec}.

For the survey data,
we see that the exponential velocity distribution model gives a marginally
better fit to the observations than the $|w|^{2/3}$ or Gaussian distributions.
There is no obvious trend of \wrms\ with separation $\sigma$,
the 95\% confidence range for \wrms\ is roughly 100--1100 \kms,
with a maximum likelihood value around 500 \kms.
Note that our survey does not provide a strong constraint on small scale
peculiar velocities simply due to the sparse-sampling strategy we have 
employed.
It would be interesting to compare $f(w)$ for the morphological and
luminosity-selected subsamples defined in Paper~2, but
unfortunately, the shot-noise errors on $\xi(\sigma,\pi)$ for these 
subsamples are 
too large to test for variation of \wrms\ with galaxy type or luminosity.

For both sets of simulations, the exponential model provides the best
fit for the smallest $\sigma$ bin ($\sigma < 2 \hMpc$) but this model
fairs less well at larger projected separation.
In fact none of the models for $f(w)$ provides a very good fit to the 
simulations, which probably explains why \wrms\ is being overestimated
for the simulations at small $\sigma$.
(Using the position and velocity
information for each galaxy in the simulations yields direct measurement of
rms peculiar velocities of 
350 km/s and 510 km/s for the LCDM and MDM model respectively.)
Our fits to \wrms\ in Table~\ref{tab:v_pec} are giving close to the
expected values for $\sigma \approx 5 \hMpc$.
We conclude that the LCDM model has slightly smaller peculiar velocities 
than the data,
whereas peculiar velocities in the MDM model are comparable to those
measured from the Stromlo-APM data.

\section{Testing Estimators for $\beta$}
\label{sec:tests}

In the preceding section we showed that the simulations have comparable
two-point clustering statistics and small-scale
peculiar velocities to the survey data.
One might naively expect that if the simulations obey linear
theory at a certain scale then linear theory should also be applicable to the
survey data on that same scale.
However, as seen below, we find that the MDM simulations show evidence for
non-linear effects to much larger scales than the LCDM simulations,
despite the relatively small difference in small-scale peculiar velocities.
As pointed out by \refer{Fisher and Nusser (1996)}, the departure from
linear theory is due to non-linear streaming (as modelled, for example,
by the Zel'dovich approximation), and not just due to virialized motions.

In this section we investigate the scales on which linear theory is
obeyed by the simulations, insofar as one can estimate $\beta$ from
the {\em anisotropy} of clustering in redshift space and from the
{\em amplification} in $\xi(s)$ in redshift space compared to real space.

\subsection{Anisotropy of $\xi(\sigma,\pi)$}
\label{sec:xi_l}

We estimate the anisotropy in $\xi(\sigma,\pi)$ as follows.
The redshift space spherical harmonics $\xi_l(r)$ in 
equation~(\ref{eqn:beta_02}) are given by an integral
over the full redshift space correlation function $\xi(r,\mu)$, 
\be
  \xi_l(r) = \frac{2l + 1}{2}\int_{-1}^1 \xi(r,\mu) P_l(\mu) d\mu,
\ee
where $P_l(\mu)$ is the $l$th order Legendre polynomial.
We can measure $\xi(r,\mu)$ by comparing the observed, weighted sum of galaxy
pairs $w_{dd}(r,\mu)$ at separation $r$ and direction cosine to line of 
sight $\mu$, with the expected background bgr($r,\mu$)
for an isotropic, unclustered distribution.
The line of sight direction is defined as the bisector in angle of each pair.

Since $P_l(\mu)$ is an odd function for odd $l$ and an even function for
even $l$, the odd-$l$ harmonics vanish and the even-$l$ harmonics are given
by
\begin{eqnarray}
  \xi_l(r) &=& (2l+1)\int_0^1 \left(\frac{w_{gg}(r,\mu)}{{\rm bgr}(r,\mu)} -1
   \right) P_l(\mu) d\mu \nonumber\\
   &\approx & (2l+1) \Delta \mu \left[\sum_{\rm pairs(r)} 
    \frac{w_i w_j P_l(\mu)}{{\rm bgr}(r,\mu)}\right] - \delta_{l0}.
  \label{eqn:xi_l}
\end{eqnarray}
Here we have replaced the integral over $\xi(r,\mu)$ with respect to $\mu$
by the weighted and appropriately normalised sum of $P_l(\mu)$ for all
galaxy pairs at separation $r$.
The $w_i w_j$ are the products of the weights of each galaxy
in the pair, given by equation (1) of Paper~2 and
$\delta_{l0}$ is the Kronecker delta symbol, equal to unity for $l=0$,
zero otherwise.
The background bgr($r,\mu$) is obtained by linear interpolation in $\mu$ from
a pre-calculated look-up table.
We do not interpolate between $r$-bins, since the $\xi_l(r)$ are calculated
in the same separation bins in which bgr($r,\mu$) is tabulated.
This look-up table is generated in ten fixed steps in $\Delta \log r$
and $\Delta \mu$ using a large catalogue of
random points with the same selection function and within the same boundaries
as the data, and using the same weighting scheme,
\be
  {\rm bgr}(r,\mu) = \frac{[w_{gr}(r,\mu)]^2}{w_{rr}(r,\mu)}.
  \label{eqn:bgr}
\ee
Here $w_{gr}(r,\mu)$ and $w_{rr}(r,\mu)$ are the summed weights of 
galaxy-random and random-random pairs respectively.
This definition of the background does not require one to estimate
the mean density of the data and random catalogues to normalize $\xi$;
instead the normalisation is determined using only those galaxies at
separation $r$ and cosine direction $\mu$.
Such an estimator gives more stable estimates of $\xi$ on
large scales than traditional estimators (Hamilton 1993b; Paper~2).
In Figures~\ref{fig:xi_l_nsim}a and b we plot the $\xi_l(r)$ measured from the 
LCDM and MDM simulations respectively as the points with error bars.
Also shown by the curves are the linear-theory predictions for the $\xi_l(r)$
using equations (6)--(9) of Hamilton (1992).

For the LCDM model (Figure~\ref{fig:xi_l_nsim}a), we use the power spectrum of 
\refer{Efstathiou, Bond and White (1992)} with $\Gamma = 0.2$
and $\beta=0.38$.
In this theory, $\Gamma = \Omega_0 h$ determines the shape of the real-space
correlation function, and $\beta$ determines the redshift-space distortions.
Since the simulations are unbiased, $\beta = \Omega_0^{0.6} = 0.38$.
We see that the direction-averaged correlation function $\xi_0$
measured from the simulations agrees well with linear theory on scales
as small as $2 \hMpc$.
Non-linearity is a far more severe problem for the quadrupole ($\xi_2$)
and hexadecapole ($\xi_4$) harmonics.
The quadrupole harmonic is expected to be negative in the linear regime
(Hamilton 1992) and so we plot $-\xi_2$ in Figure~\ref{fig:xi_l_nsim}.
The measured $\xi_2$ is in fact positive on scales $r \simlt 15 \hMpc$
suggesting that non-linear effects dominate $\xi_2$ on
these scales.
The amplitude of $\xi_2$ is lower than that predicted by linear theory
until $r \approx 30 \hMpc$, suggesting that non-linearity may affect
$\xi_2$ out to these scales.
Beyond $\sim 55 \hMpc$ the data is too noisy to measure $\xi_2$ reliably.
The hexadecapole harmonic $\xi_4$ has an
amplitude on scales $r \simlt 15 \hMpc$ exceeding the linear theory 
prediction by several orders
of magnitude (in fact it is comparable to the direction-averaged correlation
$\xi_0$) and on larger scales, it's measurement is too noisy to be useful.
Equation~(\ref{eqn:beta_02}) gives physically reasonable estimates of
the quantity $\beta$ over a range of scales 21--42 $\hMpc$:
$\beta = 0.06 \pm 0.43$, $0.47 \pm 0.48$ and $0.02 \pm 0.35$
respectively at separations of $r \approx 21$, 30 and 42 $\hMpc$.
Comparison with the linear theory curves suggests that
the first of these estimates (at $r \approx 21 \hMpc$) is probably
biased low by non-linearity, but the second two estimates appear to
be in the linear regime.

For the MDM model (Figure~\ref{fig:xi_l_nsim}b), 
we use the power spectrum of \refer{Klypin \etal\ (1993)}.
Again, we see that the direction-averaged correlation function $\xi_0$
agrees very well with linear theory over all scales measured but that
$\xi_2$ is only in reasonable agreement for scales $r \simgt 15 \hMpc$.
The measured errors for $\xi_4$ are very large on all scales, so we
do not show the $\xi_4$ estimates on this plot.
At scales $r \approx 21$, 30 and 42 $\hMpc$, equation~(\ref{eqn:beta_02})
provides estimates of $\beta$ of $0.30 \pm 0.31$, $0.39 \pm 0.44$
and $0.61 \pm 0.86$, all of which underestimate the true value $\beta = 1$.

\subsection{Amplification of $\xi(s)$}
\label{sec:nsim_sr}

We have seen in the previous subsection that {\em anisotropies} in
the redshift space correlation function only agree with the linear
theory prediction on very large scales, where measurements from
existing galaxy catalogues are too noisy to usefully constrain $\beta$.
In this subsection, we investigate how well one might measure $\beta$
from the {\em amplification} of the direction-averaged correlation
function in redshift space compared to real space.

The ratio $\xi(s)/\xi(r)$ is subject to large random fluctuations in
the linear regime where $\xi(r)$ is small.
Therefore, when estimating $\beta$ from equation (\ref{eqn:beta_sr})
it is desirable to either fit to $\beta$ measured over a range of scales
or alternatively to smooth the $\xi$ estimates before taking their ratio.
A common way of smoothing $\xi(r)$ is to take its volume integral
\be
  J_3(r) = \int_0^r x^2 \xi(x) dx.  \label{eqn:J3}
\ee

With an $N$-body simulation, one has the advantage of being able to estimate
the real space correlation function directly by using the real space locations
of the simulation galaxies in (\ref{eqn:xi_ni}), as well as via inversion of 
the projected correlation function (\ref{eqn:xi_proj}).
The former estimate of $\xi(r)$ is useful for studying effects of
non-linearity in the simulations, whereas the latter, noisier, estimate gives
a more realistic assessment of what we can hope to measure from real data.
We have used the ratios $\xi(s)/\xi(r)$ and $J_3(s)/J_3(r)$
measured with both $\xi(r)$
estimates in equation~(\ref{eqn:beta_sr}) to estimate $\beta$ on a range of 
scales.
These estimates are presented in Figures~\ref{fig:beta_lcdm} and 
\ref{fig:beta_mdm} for the LCDM and MDM simulations respectively.

For the LCDM simulations (Fig.~\ref{fig:beta_lcdm}), we see that the
ratio $\xi(s)/\xi(r)$ (open symbols)
appears to converge to give the correct value of $\beta$
by a scale $r \approx 5 \hMpc$, although the estimates are biased a little
high.
Beyond $r \approx 20 \hMpc$, the measured correlation functions are
too noisy to usefully constrain $\beta$.
We have calculated a maximum-likelihood fit to the four data points between
5 and 16 $\hMpc$, and find that $\beta = 0.56$ with a 95\% confidence
range 0.44--0.68.
The ratio $J_3(s)/J_3(r)$ (solid symbols) is biased low by non-linear
dynamics to larger scales ($r \approx 20 \hMpc$), but by these scales
is providing an almost unbiased and low-noise estimate of the true $\beta$.
The point at $r = 17.8 \hMpc$ yields $\beta = 0.29 \pm 0.07$ (one sigma error).
The results using $\xi(r)$ and $J_3(r)$ inferred from inversion
of the projected correlation function (Fig.~\ref{fig:beta_lcdm}b)
are only a little noisier than those determined from the direct estimates
of $\xi(r)$ and $J_3(r)$.
Fitting to $\beta_\xi$ over the same four measurements provides
$\beta = 0.58$ with a 95\% confidence limit 0.41--0.76 and
$\beta_{J_3}(17.8) = 0.36 \pm 0.13$.

For the MDM simulations (Fig.~\ref{fig:beta_mdm}), we see that the
ratio $\xi(s)/\xi(r)$ (open symbols) does not converge so readily
to the correct value of $\beta$; it crosses $\beta = 1$ at $r \approx 10 \hMpc$
but then tends to overshoot slightly.
Again, beyond $r \approx 20 \hMpc$, the measured correlation functions are
too noisy to usefully constrain $\beta$.
A fit to $\beta_\xi$ using the two measurements in the range 10--16 $\hMpc$
provides the estimates $\beta = 1.41$ (1.00--1.83) and
$\beta = 0.93$ (0.72--1.15) for the direct and inverted estimates
of $\xi(r)$ respectively.
The ratio $J_3(s)/J_3(r)$ (solid symbols) converges to $\beta = 1$ by scales
$r \approx 20 \hMpc$; estimates on larger scales tend to overestimate
$\beta$, but most are within $\sim 1 \sigma$ of $\beta = 1$.
We measure $\beta_{J_3}(17.8) = 1.02 \pm 0.16$ and 
$\beta_{J_3}(17.8) = 0.85 \pm 0.17$ from Figs.~\ref{fig:beta_mdm}
(a) and (b) respectively.

One could consider using alternative, smoothed measures of the galaxy 
clustering which are less affected by small-scale peculiar velocities 
than $J_3$.
An example would be a modified form of the $J_3$ integral in which the
lower limit of integration is set at, say, $r_{\rm min} = 5 \hMpc$.
This was tried, and while the contribution from peculiar velocities
is indeed suppressed, the modified estimator is subject to larger
random fluctuations than $J_3$ as defined in (\ref{eqn:J3}).

\subsection{Conclusions from Simulations}

We have analysed two ensembles of CDM-like $N$-body simulations,
one a low density (LCDM, $\Omega_0 = 0.2$) model, and one high density
(MDM, $\Omega_0 = 1$).
Both give a reasonable match to the real and redshift space
correlation functions measured from the Stromlo-APM Redshift Survey
and have comparable peculiar velocities $\wrms$.
We used these simulations to investigate the practicality of measuring
the quantity $\beta = \Omega_o^{0.6}/b$ from (a) the anisotropy and
(b) the amplification of the measured galaxy clustering in redshift space.

For the LCDM simulations,
we find that the 2nd and 4th order spherical harmonics of the redshift
space correlation function are severely affected by non-linearities
to scales $r \sim 30 \hMpc$, and so one needs a reliable measurement
of the $\xi_l$ on scales $r \simgt 30 \hMpc$ in order for linear theory
to be applicable.
This is in agreement with the Fourier-space harmonic analysis of
Cole \etal\ (1994).
The situation is even worse for the MDM simulations, where the
2nd order spherical harmonic underestimates $\beta$ on {\em all} scales
at which it can be measured reliably.

By contrast,
the amplitude of the direction averaged correlation function $\xi_0$
is relatively weakly affected by non-linearity.
For the LCDM model
non-linear effects become unimportant on much smaller scales, 
$r \sim 5 \hMpc$ for $\xi$ measures and $r \sim 20 \hMpc$ for $J_3$ measures.
The MDM simulations effectively agree with linear theory by scales of
$r \approx 10 \hMpc$ for the $\xi$ measures and by
$r \approx 20 \hMpc$ for the $J_3$ measures.

Thus the more practical method for determining $\beta$ from current
redshift surveys is by comparison of the direction-averaged redshift-space
correlation function with the real-space correlation function.
By using the $J_3$ volume integral over $\xi$, one decreases the
random noise in $\beta$, at the expense of pushing out the effects of
non-linearity to larger scales.
In practice it will be worthwhile to estimate $\beta$ from both 
$\xi(s)/\xi(r)$ and $J_3(s)/J_3(r)$.
We expect the $J_3$ ratio to give the less noisy results.

\section{Results from Stromlo-APM Survey}
\label{sec:results}

\subsection{Constraints on $\beta$}

We have measured the spherical harmonics $\xi_l(s)$ of the Stromlo-APM
redshift space correlation function in the same way as for the simulations
(\S\ref{sec:xi_l}).
The results are shown as the points with
error bars in Figure~\ref{fig:xi_l_apm}.
The curves show the linear theory predictions for LCDM (light lines)
and MDM (heavy lines).
We see pronounced effects of non-linearity and very large error bars
on the $\xi_2$ harmonics on scales smaller than $20 \hMpc$.
The direction-averaged correlation function $\xi_0$ is well matched
by the LCDM linear theory prediction on all scales.
The MDM model slightly under-predicts $\xi_0$ on small and large scales
(cf Fig.~\ref{fig:xi_apm_nsim}).
The measured $\xi_2$ harmonics agree quite well with the LCDM linear
theory prediction on scales 20--50 $\hMpc$ whereas the MDM prediction
is too high.
In other words, we see evidence for smaller redshift space distortions 
than expected in an unbiased $\Omega_0 = 1$ model.
On scales of 21, 30 and 42 $\hMpc$, the estimated values of $\beta$
are $0.41 \pm 0.17$, $-0.03 \pm 0.29$ and $0.23 \pm 0.31$ respectively.

From our analysis of simulated galaxy catalogues in the previous section,
we expect the ratios of the direction-averaged clustering
in redshift space and real space to give more reliable estimates of $\beta$.
In Figure~\ref{fig:beta_apm}, we plot our estimates of $\beta$ 
from the redshift survey, as determined from the ratio $\xi(s)/\xi(r)$
(open symbols) and from the ratio $J_3(s)/J_3(r)$ (solid symbols).
The error bars in this figure are determined from the scatter between
estimates of $\beta$ for each of the nine bootstrap resampled surveys,
and are in reasonable agreement with the scatter found between different
realizations of the simulations (Figs.~\ref{fig:beta_lcdm} and
\ref{fig:beta_mdm}) in the linear regime.
There is a wide scatter in the $\beta$ estimates from the $\xi$ ratios
around 10--20 $\hMpc$, but the $J_3$ ratios give consistent results
over this range, appearing to favour $\beta \approx 0.2$--0.5.
Even the point giving the largest value of $\beta$ ($\beta = 0.48 \pm 0.12$ at 
$r = 17.8 \hMpc$) favours $\beta \ll 1$.
It is interesting to note that in both sets of simulations, 
the $J_3$ ratio at this scale provides an estimate of $\beta$ within 1 sigma
of the correct value.
Therefore $\beta \simlt 0.6$ would seem to be a reasonable upper limit allowed
by our analysis and we conclude that either $\Omega_0 <1$ or that galaxies
are significantly biased tracers of the mass distribution, $b \simgt 2$.

As an aside, we note that the small-scale, non-linear, behaviour of $\beta$ for
the Stromlo-APM survey is markedly different to that of the simulations,
which have a systematically negative $\beta$ and smaller errors on
scales $r \simlt 5 \hMpc$.
We are not overly concerned by this as the simulations provide a relatively
poor fit to observed clustering on small scales and we know that
equation (\ref{eqn:beta_sr}) is clearly not to be trusted in regimes in which
it predicts a negative $\beta$!

\subsection{Disentangling $\Omega$ and $b$}

Ideally, of course, one would like to know the values of the density 
parameter $\Omega$ and the bias parameter $b$ individually.
Cole \etal\ (1994) have discussed how it may be possible with future
redshift surveys to separate the determination of $\Omega$ and $b$ 
by studying the scaling of non-linear effects.
An alternative approach is the study of high order correlations to constrain 
the (possibly non-linear) biasing model using weakly non-linear perturbation 
theory.
Gazta\~{n}aga and Frieman (1994)\refeq{Gaztanaga and Frieman 1994}
have used high order moments
of APM galaxies to constrain biasing models.
To be consistent with non-linear perturbation theory,
one should allow the possibility of {\em non-linear}
bias, in which case second- and third-order non-linear bias coefficients 
can be chosen which match the observed high order correlations.
However, the observations are very well fit by
an {\em unbiased}, low-density CDM model with $b \approx 1$.
Applying Occam's razor, this seems the more natural solution.

\subsection{Relative Biasing of Different Galaxy Types}
\label{sec:bias}

Despite the uncertainties in the value of galaxy bias with respect to the mass,
one may study the {\em relative} bias of galaxies of
different type by comparing their clustering properties.
There are two independent ways to measure the relative bias factors:
(1) by comparing real-space clustering in the linear regime and 
(2) by measuring the redshift space amplification of clustering due to 
equation~(\ref{eqn:beta_sr}).
One thus has a consistency check on measurements of $\beta$ and relative
bias parameters.
We have already compared the clustering of galaxies of different luminosity
and morphological type in Paper~2.
In that paper we concentrated on the small-scale (power-law) regime of
galaxy clustering.
Here we study clustering in the linear regime.
We analyse galaxy subsamples from the Stromlo-APM
survey of low, middle and high luminosity, and of early and late type.
These galaxy samples are defined in Table~1 of Paper~2.

In Figure~\ref{fig:bias} we plot the relative bias values 
$b_\xi = \xi_{tg}(r)/\xi_{gg}(r)$ (open symbols) and
$b_{J_3} = {J_3}^{tg}(r)/{J_3}^{gg}(r)$ (solid symbols) for each
sample over a range of scales.
We have divided the real-space cross-correlation of each sample
with the parent APM galaxy sample (Paper~2) by the real-space
cross-correlation function of the `all galaxies' sample, 
hence the relative bias
for the `all galaxies' sample is defined to be unity.
The error bars are determined from the one sigma scatter in $b_\xi$ from
the nine bootstrap-resampled versions of the survey data.
Given that our estimates of $\xi(r)$ are unreliable beyond 
$r \approx 20 \hMpc$, we see no obvious trend of relative bias with
scale in the linear regime.
We have therefore calculated the maximum-likelihood value of $b_\xi$ over
the range 5--12 $\hMpc$, a regime over which the assumption of linear
bias seems reasonable and where our measurement errors are not too large.
These values, along with the 95\% confidence limits, are given in
Table~\ref{tab:samples}.
We also show the relative bias $b_{J_3}$ measured at a separation of 
$r = 17.8\hMpc$ in this table.
We see that low-luminosity galaxies (sample b), are only about one third
as strongly clustered as middle-luminosity galaxies, in accordance with
the findings of Paper~2.
High luminosity galaxies (sample d) appear to be about thirty percent
more strongly clustered than middle-luminosity galaxies, although the
95\% confidence limits do allow for no difference in clustering, again
in accord with Paper~2.
Early type galaxies (e) have a very similar bias parameter to luminous 
galaxies,
whereas late type galaxies have a bias parameter midway between that of
low and middle-luminosity galaxies.

The parameter most commonly used for normalizing the power spectrum of
clustering models and theories to observations
is the variance of galaxy counts in $8 \hMpc$ radius spheres, $(\sigma_8^2)_g$,
\be
  (\sigma_8^2)_g = \frac{1}{V^2} \int\int_V \xi(r_{12}) dV_1 dV_2.
\ee
We estimate $(\sigma_8^2)_g$ by Monte-Carlo integration of the observed 
real-space correlation function for each of the galaxy samples
using 500,000 randomly placed pairs of 
points inside an $8 \hMpc$ radius sphere.
Results are given in Table~\ref{tab:samples}.
As we see from Figures~\ref{fig:beta_lcdm} and \ref{fig:beta_mdm}, 
$(\sigma_8^2)_g$ may be mildly
affected by non-linearity, and so we do not expect $b_\xi$ and $b_{J_3}$
to be directly proportional to $(\sigma_8)_g$.
Note the significant difference in $(\sigma_8^2)_g$ for galaxies fainter
than $L^*$ and those around $L^*$ and brighter.

In Figure~\ref{fig:beta_samp} we plot estimates of $\beta_\xi$ (open symbols)
and $\beta_{J_3}$ (solid symbols) for each 
galaxy sample.
In order to estimate $\beta$ for galaxy sub-samples, we have measured the
redshift-space cross-correlation function of the given galaxy sample
with the all-galaxies sample using the estimator
\be
  1 + \xi_{tg}(s) = \frac{w_{tg}(s) w_{rr}(s)}{w_{tr}(s) w_{gr}(s)}.
  \label{eqn:xix}
\ee
Here the subscript $t$ denotes galaxies of specific type,
$g$ denotes all galaxies and $r$ denotes random points 
(cf.\ eq.~\ref{eqn:xi_ni}).
Comparing this with the real-space cross correlation function $\xi_{tg}(r)$
gives an estimate of $\beta$ via equation~(\ref{eqn:beta_sr}).
Once again, the error bars come from the scatter between bootstrap resamplings.
Estimates of $\beta_\xi$ and 95\% confidence limits over the separation range
5--12 $\hMpc$ are given in Table~\ref{tab:samples}.
Clearly, negative values or lower limits on $\beta$ are not physically
reasonable, but the upper limits are still useful.
For instance, for bright galaxies (d) we can say that $\beta < 0.9$ with 95\%
confidence.
The value of $\beta_{J_3}$ at $r \approx 17.8 \hMpc$ is also given in 
Table~\ref{tab:samples}.

Now, if the `true' bias factor $b_t^{\rm true}$ 
(ie. the bias with respect to the mass)
is related to the relative bias factor by $b_t^{\rm true} = b_0 b_t$,
(so that $b_0$ is the bias factor for the `all galaxies' sample)
then the product $b_t \beta$ for each galaxy sample should be a constant
equal to $\Omega^{0.6}/b_0$.
While we do find some scatter in this quantity between the different samples,
the results are consistent ($\Omega^{0.6}/b_0 \approx 0.2$--0.6)
within the 95\% confidence limits.


\section{Conclusions}
\label{sec:concs}

We have used large and realistic $N$-body simulations to investigate the
effects of non-linearity on two estimators for the quantity $\beta$
(eqs.~[\ref{eqn:beta_02}] and [\ref{eqn:beta_sr}]) in low and high density
models.
We find that non-linearity affects the 2nd order spherical
harmonic $\xi_2(s)$ of redshift space clustering to scales as large as
$30 \hMpc$.
In contrast, the amplitude of the direction-averaged correlation function 
$\xi_0(s)$
is only weakly affected by non-linearity on scales $r \simgt 5 \hMpc$
in the LCDM model and $r \simgt 10 \hMpc$ in the MDM model.
Therefore the most practical method for constraining $\beta$ with
existing redshift surveys is by measuring the amplification of
direction-averaged redshift space clustering over real space clustering.
An alternative approach, modeling the non-linearity, has recently
been used by Cole \etal\ (1995) and Fisher and Nusser (1996).
These authors also find that $\beta \simlt 0.6$ from independent
data sets.

Measuring the projected cross-correlation of Stromlo-APM galaxies with
the parent APM survey galaxies enables a reliable determination of
$\xi(r)$ to scales $r \simlt 20 \hMpc$.
The ratio $\xi(s)/\xi(r)$ shows rather large scatter
on scales 5--20 $\hMpc$, but
the integral $J_3$ is less noisy than $\xi(r)$ and on a scale 
$r \approx 17.8 \hMpc$, $J_3(s)/J_3(r)$ provides the estimate 
$\beta = 0.48 \pm 0.12$.
A reasonable upper limit on $\beta$ from this analysis is
$\beta \simlt 0.6$.
Given that the $\beta_{J_3}$ estimates on either side of $r \approx 17.8 \hMpc$
are almost consistent with $\beta = 0$, we regard our analysis as
providing an {\em upper limit} of $\beta \simlt 0.6$, rather than an
actual estimate of $\beta$.
Although a little lower than estimates of $\beta$ from peculiar velocity
analyses (eg. Hudson \etal\ 1995, who find $\beta = 0.74 \pm 0.13$), 
the largely unknown systematic errors in most current estimates of $\beta$ 
means they are all fairly consistent.
See \refer{Dekel (1994)} or \refer{Strauss and Willick (1995)} for a review of 
recent measurements of $\beta$.

The Stromlo-APM survey is a powerful sample for constraining $\beta$
since the large volume probed enables us to reliably measure redshift space
galaxy clustering in the linear regime, whereas many previous analyses
have been limited to measuring $\xi(s)$ in the non-linear regime.
Cross-correlation with the fully-sampled APM galaxy survey enables us to
measure $\xi(r)$ much more accurately than using the angular correlation
function $w(\theta)$, and thus the technique of using the ratio
$\xi(s)/\xi(r)$ [or $J_3(s)/J_3(r)$] comes into its own for this survey.
The most likely source of systematic error in this analysis is in
the inversion of the projected correlation function $\Xi(\sigma)$ to 
obtain the real space correlation function $\xi(r)$.
Comparison of the $\beta$ determinations from the simulations using
$\xi(r)$ measured directly from the simulations and via inversion 
of the projected correlation function (\S\ref{sec:nsim_sr})
suggests that any such error is comparable to or smaller than
random errors.

With linear theory and 2-point clustering statistics alone one cannot
separate the contributions of $\Omega$ and $b$ in $\beta = \Omega^0.6/b$.
However, as Gazta\~{n}aga and Frieman (1994) have discussed, higher
order correlations may be used to constrain biasing models.
Their analysis of APM galaxies favours a linear bias parameter
$b \approx 1$, although to be strictly self-consistent, one should
allow for a non-linear bias model in non-linear perturbation theory,
in which case one can always match the observed skewness of APM galaxy counts
in cells by adjusting the non-linear bias parameters.
Further work is clearly required in constraining possible biasing models.

As we have seen in Paper~2, different classes of galaxy have different
clustering properties, and so not all galaxies can have exactly the same
biasing parameter.
In particular, low-luminosity galaxies are about three times less strongly
clustered than $L^*$ galaxies on large scales, and so
$b_{L^*} \approx 3 b_{<L^*}$.
An interesting test would be to see if high order clustering of low-luminosity
galaxies also predicts a lower value of bias than for $L^*$ galaxies.

In conclusion, we find that a relatively low value of $\beta$ is favoured by
the Stromlo-APM data; our results appear to rule out an unbiased, $\Omega = 1$
model.
We thus conclude that $\Omega < 1$ and/or that galaxies are positively biased,
ie. more strongly clustered than the underlying mass distribution.

Acknowledgments: we are indebted to Andrew Hamilton for invaluable e-mail
correspondence and to Peter Quinn for useful discussions during the early 
stages of this project.
We thank Albert Stebbins for pointing out an error in an earlier version
of this paper and the anonymous referee who suggested that we test
our methods on a high-density simulation.
JL acknowledges the hospitality of the Oxford Physics Department where
the final version of this paper was completed.

\clearpage
\section*{Tables}

\begin{table}[htbp]
 \begin{center}
 \caption{Power-law fits to real and redshift space clustering in the
	  survey data and simulations.}
 \vspace{0.5cm}
 \label{tab:xi_fits}
 \begin{math}
 \begin{array}{lcccc}
 \hline
 \hline
 {\rm Sample} & \gamma_s & s_0 & \gamma_r & r_0\\
 \hline
{\rm Stromlo-APM} & 1.49 \pm 0.15 & 5.9 \pm 0.4 & 1.71 \pm 0.05 & 5.1 \pm 0.2\\
{\rm LCDM} & 1.84 \pm 0.16 & 6.1 \pm 0.5 & 1.97 \pm 0.12 & 6.0 \pm 0.6\\
{\rm MDM} & 1.37 \pm 0.15 & 4.3 \pm 0.2 & 1.48 \pm 0.09 & 3.9 \pm 0.9\\
  \hline
  \hline
 \end{array}
 \end{math}
 \vspace{0.2cm}

Notes.---$\gamma_s$ and $s_0$ are the
power-law fit parameters to the correlation function in redshift space
measured over the range 1.5--30 $\hMpc$.
$\gamma_r$ and $r_0$ are the real-space power-law parameters over
the range 0.2--20 $\hMpc$.
\end{center}
\end{table}

\begin{table}[htbp]
 \begin{center}
 \caption{Constraints on galaxy peculiar velocities (\kms) for survey data
	  and simulations.}
 \vspace{0.5cm}
 \label{tab:v_pec}
 \begin{math}
 \begin{array}{rrrrrrrrrrrrr}
 \hline
 \hline
 {\rm Model} &  \multicolumn{4}{c}{\rm Gaussian} & 
		\multicolumn{4}{c}{|w|^{3/2}} & 
		\multicolumn{4}{c}{\rm Exponential}\\
 \sigma & 
   \wrms\ & \multicolumn{2}{c}{\rm 95\%\ conf} & \ln(\cal{L}) &
   \wrms\ & \multicolumn{2}{c}{\rm 95\%\ conf} & \ln(\cal{L}) &
   \wrms\ & \multicolumn{2}{c}{\rm 95\%\ conf} & \ln(\cal{L})\\
 \hline
 \multicolumn{13}{c}{\rm Survey\ Data} \\
   1.0 &  383 &   89 &  756 &-10.1 &  404 &   96 &  841 & -9.9 &  438 &  101 & 1095 & -9.9 \\
   3.0 &  400 &  203 &  800 & -3.1 &  446 &  219 &  889 & -2.9 &  531 &  253 & 1081 & -2.6 \\
   5.0 &  406 &   98 &  832 & -3.0 &  417 &  105 &  907 & -2.9 &  459 &  112 & 1093 & -2.9 \\
   7.0 &  271 &    0 &  708 &  0.0 &  271 &    0 &  796 &  0.0 &  282 &    0 & 1040 & -0.1 \\
  \hline
 \multicolumn{13}{c}{\rm LCDM\ Simulation} \\
   1.0 &  473 &  443 &  504 &-32.8 &  500 &  461 &  538 &-20.7 &  569 &  519 &  631 &-11.6 \\
   3.0 &  452 &  388 &  524 &  2.8 &  475 &  405 &  554 &  2.3 &  533 &  447 &  634 & -0.2 \\
   5.0 &  228 &  153 &  299 &  7.5 &  234 &  155 &  311 &  7.4 &  247 &  158 &  336 &  7.1 \\
   7.0 &  283 &  185 &  380 & 10.6 &  293 &  187 &  396 & 10.5 &  310 &  191 &  432 & 10.2 \\
  \hline
 \multicolumn{13}{c}{\rm MDM\ Simulation} \\
   1.0 &  628 &  560 &  726 & -0.5 &  698 &  610 &  788 &  3.6 &  847 &  743 &  975 &  5.3 \\
   3.0 &  480 &  425 &  543 & 10.3 &  513 &  453 &  585 & 10.5 &  601 &  521 &  690 &  8.0 \\
   5.0 &  472 &  395 &  566 &  8.2 &  498 &  409 &  601 &  7.2 &  552 &  448 &  691 &  4.9 \\
   7.0 &  396 &  302 &  497 &  0.3 &  404 &  305 &  515 & -0.7 &  430 &  313 &  560 & -2.8 \\
  \hline
 \end{array}
 \end{math}
 \end{center}
\end{table}

\begin{table}[htbp]
 \begin{center}
 \caption{Relative bias $b_\xi$ and $b_{J_3}$, variance in $8 \hMpc$ cells 
	  $(\sigma_8^2)_g$, 
	  and redshift space distortion factors $\beta_\xi$ and $\beta_{J_3}$ 
	  for galaxy subsamples}
 \vspace{0.5cm}
 \label{tab:samples}
 \begin{math}
 \begin{array}{clrrrrrrrrr}
 \hline
 \hline
 & {\rm Type} & \multicolumn{1}{c}{b_\xi} & 
\multicolumn{2}{c}{\rm 95\%\ conf} &
\multicolumn{1}{c}{b_{J_3}} & 
\multicolumn{1}{c}{(\sigma_8^2)_g} & 
\multicolumn{1}{c}{\beta_\xi} & 
\multicolumn{2}{c}{\rm 95\%\ conf} &
\multicolumn{1}{c}{\beta_{J_3}}\\
 \hline
 a & {\rm All} & 1.00 & 1.00 & 1.00 & 1.00 \pm 0.00 & 0.89 \pm 0.05 & 0.36 & -0.03 & 0.75 & 
0.48 \pm 0.12\\
 b & {\rm Faint} & 0.31 & 0.05 & 0.56 & 0.32 \pm 0.18 & 0.40 \pm 0.10 & 1.91 & 0.35 & 3.46 &
0.85 \pm 0.71\\
 c & {\rm Middle} & 1.05 & 0.85 & 1.26 & 1.15 \pm 0.13 & 1.08 \pm 0.14 & 0.26 & -0.12 & 0.65 &
0.59 \pm 0.29\\
 d & {\rm Bright} & 1.34 & 0.78 & 1.89 & 1.45 \pm 0.35 & 1.20 \pm 0.18 & -0.11 & -1.14 & 0.91 &
0.19 \pm 0.43\\
 e & {\rm E \& S0} & 1.39 & 0.83 & 1.95 & 1.36 \pm 0.35 & 1.24 \pm 0.27 & 0.44 & -0.13 & 0.99 &
0.27 \pm 0.59\\
 f & {\rm Sp \& Irr} & 0.78 & 0.62 & 0.95 & 0.87 \pm 0.12 & 0.66 \pm 0.05 & 0.30 & -0.16 & 0.75 &
0.33 \pm 0.45\\
  \hline
  \hline
 \end{array}
 \end{math}
\end{center}
\end{table}

\clearpage

\vspace{.5in}
\centerline{\bf REFERENCES}
\vspace{.2in}
{\refmode
 
 
Barrow, J.D., Bhavsar, S.P. and Sonoda, D.M., 1984, MNRAS, 210, 19p
 
Bean, A.J., Efstathiou, G., Ellis, R.S., Peterson, B.A.
and Shanks, T., 1983, MNRAS, 205, 605
 
Cole, S., Fisher, K.B. and Weinberg, D.H., 1994, MNRAS, 267, 785
 
Cole, S., Fisher, K.B. and Weinberg, D.H., 1995, MNRAS, 275, 515
 
Croft, R.A.C. and Efstathiou, G., 1994, MNRAS, 268, L23
 
Davis, M. and Peebles, P.J.E., 1983, ApJ, 267, 465

Dekel, A., 1994, ARA\&A, 32, 371
 
Efstathiou, G., Bond, J.R. and White, S.D.M., 1992, MNRAS, 258, 1p
 
Efstathiou, G., Ellis, R.S. and Peterson, B.A., 1988, MNRAS, 232, 431 (EEP)
 
Fisher, K.B., Davis, M., Strauss, M.A., Yahil, A. and Huchra, J.P., 1994a,
MNRAS, 267, 927

Fisher, K.B. and Nusser, A., 1996, MNRAS, in press
 
Fisher, K.B., Scharf, C.A. and Lahav, O., 1994b, MNRAS, 266, 219
 
Gazta\~{n}aga, E. and Frieman, J.A., 1994, ApJ, 437, L13
 
Hamilton, A.J.S., 1992, ApJ, 385, L5
 
Hamilton, A.J.S., 1993a, ApJ, 406, L47
 
Hamilton, A.J.S., 1993b, ApJ, 417, 19
 
Heavens, A.F. and Taylor, A.N., 1995, MNRAS, 275, 483

Hudson, M., Dekel, A., Courteau, S., Faber, S.M. and Willick, J.A., 1995,
MNRAS, 274, 305
 
Kaiser, N., 1987, MNRAS, 227, 1
 
Loveday, J., Efstathiou, G., Peterson, B.A. and Maddox, S.J., 1992a,
ApJ, 400, L43
 
Loveday, J., Peterson, B.A., Efstathiou, G. and Maddox, S.J., 1992b, ApJ,
390, 338 (Paper~1)
 
Loveday, J., Maddox, S.J., Efstathiou, G., and Peterson, B.A., 1995,
ApJ, 442, 457 (Paper~2)
 
Loveday, J., Peterson, B.A., Maddox, S.J. and Efstathiou, G., 1996,
submitted to ApJ
 
Maddox, S.J., Sutherland, W.J. Efstathiou, G., and Loveday, J.,
1990a, MNRAS, 243, 692
 
Maddox, S.J., Efstathiou, G. and Sutherland, W.J., 1990b, MNRAS, 246, 433
 
Saunders, W., Rowan-Robinson, M. and Lawrence, A., 1992, MNRAS, 258, 134
 
Strauss, M.A. and Willick, J.A., 1995, Phys.\ Rep., 261, 271
 
Tadros, H., and Efstathiou, G., 1996, in preparation
 
}  

\newpage
\section*{Figure Captions}

\begin{description}
\refstepcounter{figure}\label{fig:xi_cpts}
\item[Figure \thefigure] A contour plot of the full redshift space 
	correlation function
	$\xi(\sigma,\pi)$ measured from the Stromlo-APM Redshift Survey
	as a function of separation parallel ($\pi$) and 
	perpendicular ($\sigma$) to the line of sight
	and smoothed with a
	{\small 
	$\left\{ \begin{array}{ccc}
	1 & 2 & 1\\
	2 & 4 & 2\\
	1 & 2 & 1
	\end{array}\right\} $}
	smoothing filter.
	The contours are plotted in fixed steps in $\log \xi$ from $-3$ to 1.
	Solid contours show values $\xi \ge 1$, dashed contours show values
	$\xi < 1$ (\ie in the linear regime).

\refstepcounter{figure}\label{fig:xi_apm_nsim}
\item[Figure \thefigure] Comparison of the Stromlo-APM (filled circles),
	LCDM (open circles) and MDM (open squares) correlation functions 
	in (a) redshift space
	and (b) in real space.

\refstepcounter{figure}\label{fig:v_pec}
\item[Figure \thefigure] The redshift-space correlation function 
	$\xi(\sigma,\pi)$ plotted as a function of separation $\pi$
	along the line of sight for four bins in projected separation
	$\sigma$.
	The points with error bars show $\xi(\sigma,\pi)$ calculated
	from (a) the Stromlo-APM survey, (b) the LCDM simulation and
	(c) the MDM simulation.
	The curves show predictions for three peculiar velocity
	distribution functions --- see text for details.

\refstepcounter{figure}\label{fig:xi_l_nsim}
\item[Figure \thefigure] (a) The First three spherical harmonics for the 
	redshift-space correlation function measured from the LCDM simulations
	(symbols) and predicted by linear 
	CDM theory (curves).
	Filled circles and the continuous line shows the direction-averaged
	correlation function ($\xi_0$).
	Star symbols and the dashed line show the negative of the quadrupole
	harmonic ($-\xi_2$).
	Open circles and the dotted line show the hexadecapole harmonic
	($\xi_4$).
	(b) As (a) for the MDM simulations.  $\xi_4$ is very noisy for
	this set of simulations and so is not shown here.

\refstepcounter{figure}\label{fig:beta_lcdm}
\item[Figure \thefigure] Estimates of $\beta$ as a function of separation
	from the LCDM simulations using equation~(\ref{eqn:beta_sr}),
	(a) uses `direct' estimation of $\xi(s)$ and $\xi(r)$,
	(b) uses the projected cross-correlation estimate of $\xi(r)$
	(Eq.~\ref{eqn:xi_proj}).
	The open symbols show $\beta$ estimated from the ratio
	$\xi(s)/\xi(r)$ and the solid symbols show $\beta$ estimated 
	from the ratio $J_3(s)/J_3(r)$.
	The horizontal unbroken line shows the maximum-likelihood fit to 
	$\beta_\xi$ over the range indicated and
	the dotted lines show 95\% confidence limits on this $\beta$ estimate.
	The horizontal dashed line indicates the actual $\beta = 0.38$
	for these simulations.

\refstepcounter{figure}\label{fig:beta_mdm}
\item[Figure \thefigure] As Figure~\ref{fig:beta_lcdm} but for the
	MDM simulations.

\refstepcounter{figure}\label{fig:xi_l_apm}
\item[Figure \thefigure] The 0th and 2nd order spherical harmonics of the 
	redshift-space correlation function measured from the Stromlo-APM
	survey.  The curves are from LCDM linear theory
	(light lines) and MDM theory (heavy lines).

\refstepcounter{figure}\label{fig:beta_apm}
\item[Figure \thefigure] Estimates of $\beta$ as a function of separation
	for the Stromlo-APM survey data from the ratio $\xi(s)/\xi(r)$ 
	(open circles) and $J_3(s)/J_3(r)$ (filled circles).

\refstepcounter{figure}\label{fig:bias}
\item[Figure \thefigure] Relative bias factors for subsamples of the
	Stromlo-APM Survey, determined from the ratio 
	$\xi_{tg}(r)/\xi_{gg}(r)$ (open circles) and
	${J_3}^{tg}(r)/{J_3}^{gg}(r)$ (filled circles).
	The horizontal lines show the maximum likelihood fit to 
	$\xi_{tg}(r)/\xi_{gg}(r)$ over the range 5--12 $\hMpc$;
	the dashed lines show the 95\% confidence limits.

\refstepcounter{figure}\label{fig:beta_samp}
\item[Figure \thefigure] Estimates of $\beta$ for subsamples of the
	Stromlo-APM Survey, determined from the ratio 
	$\xi_{tg}(s)/\xi_{tg}(r)$ (open circles) and
	${J_3}^{tg}(s)/{J_3}^{tg}(r)$ (filled circles).

\end{description}

\epsfxsize=0.95\textwidth
\setcounter{figure}{0}

\begin{figure}[p]
\epsfxsize=0.95\textwidth
\epsfbox{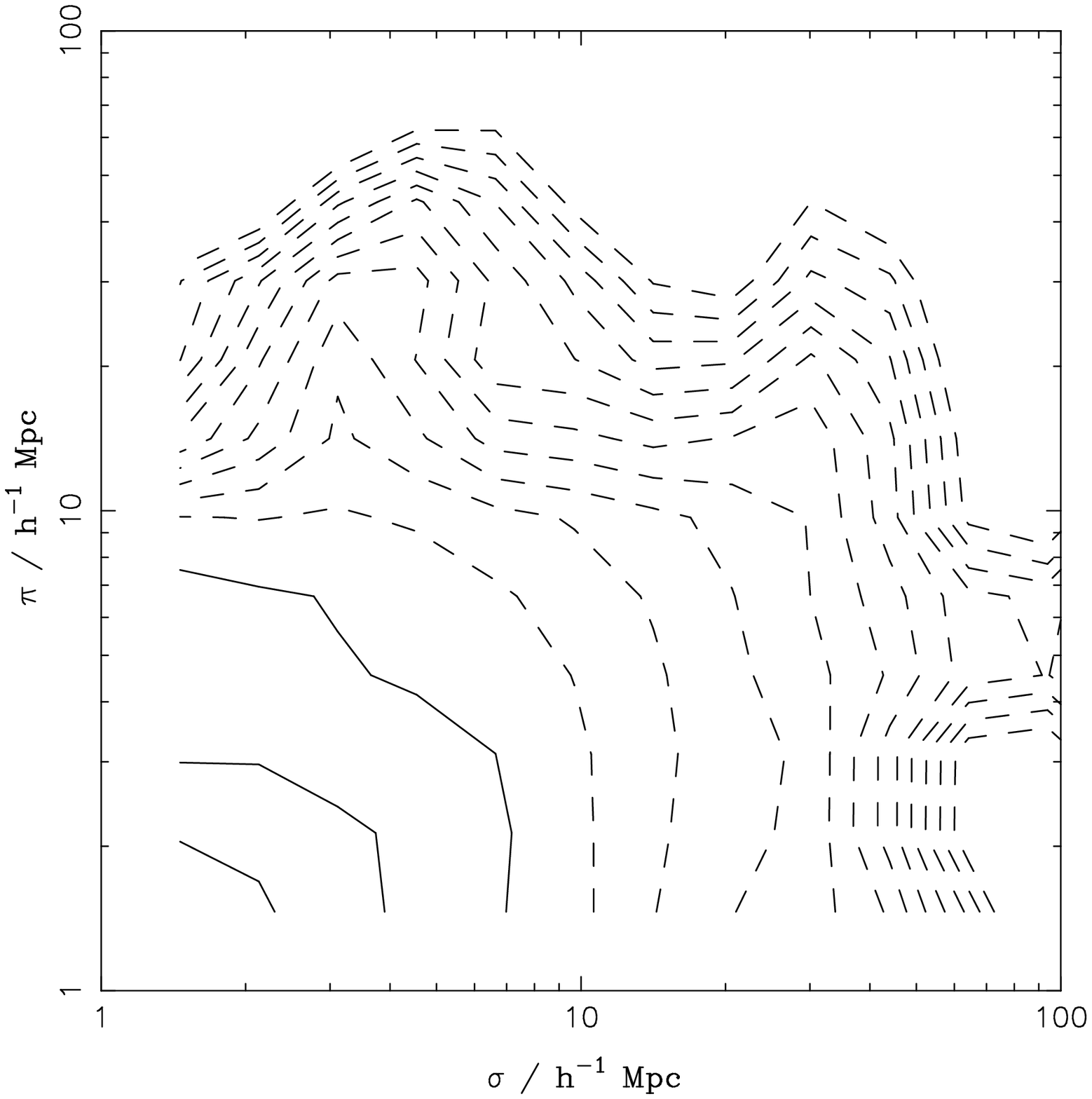}
\caption{\relax}
\end{figure}

\begin{figure}[p]
\epsfxsize=0.95\textwidth
\epsfbox{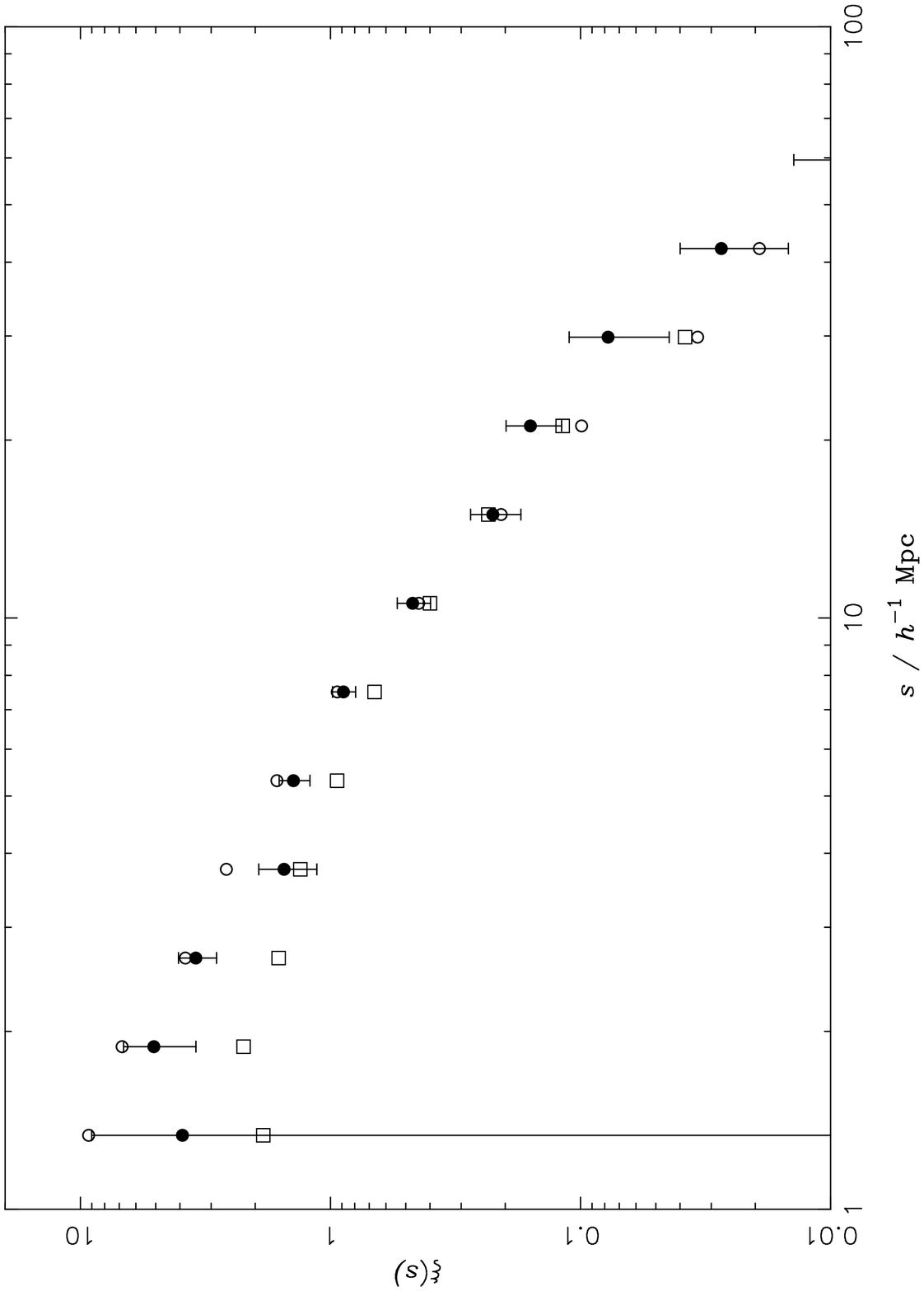}
\caption{a}
\end{figure}

\addtocounter{figure}{-1}
\begin{figure}[p]
\epsfxsize=0.95\textwidth
\epsfbox{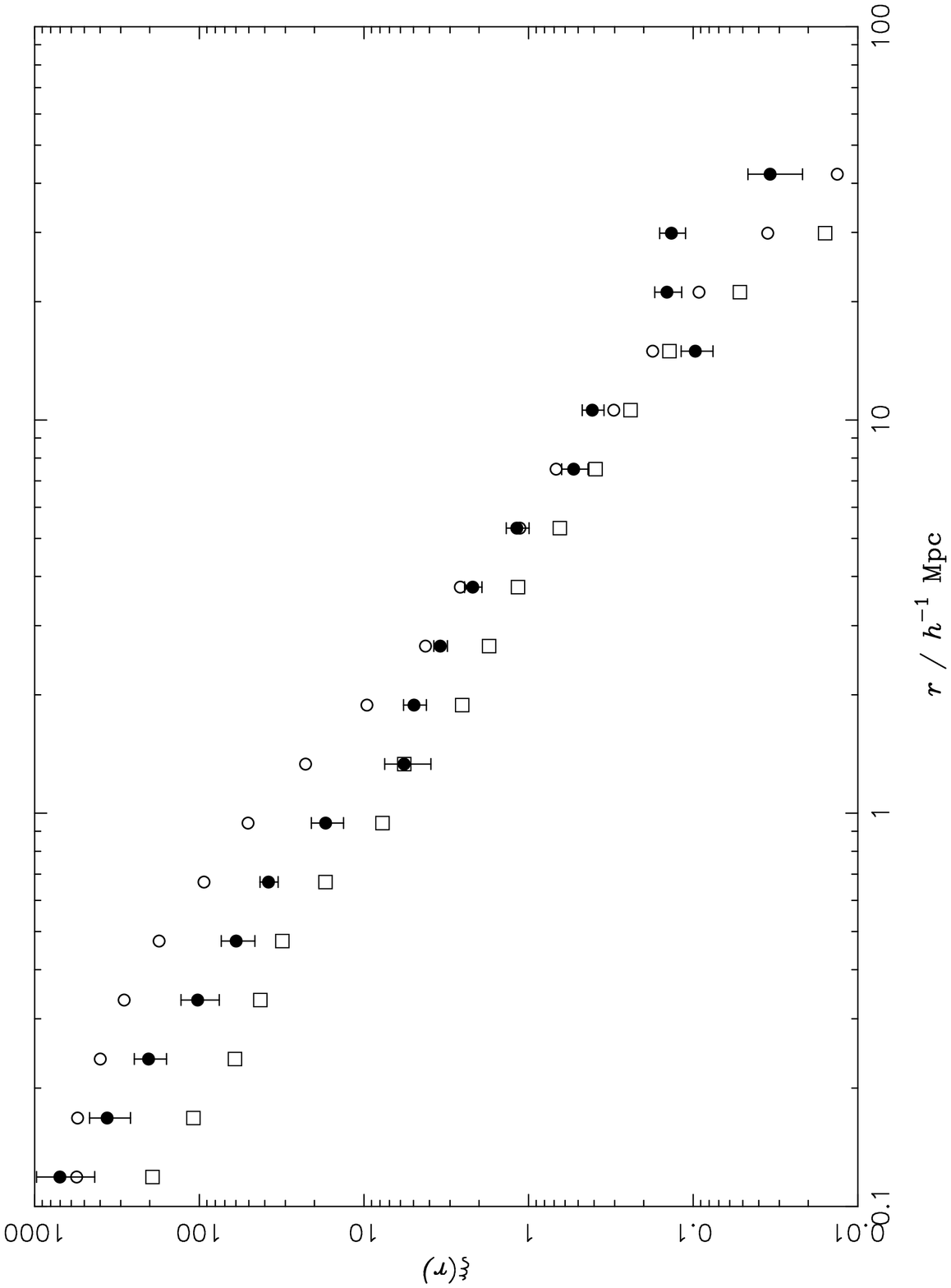}
\caption{b}
\end{figure}

\begin{figure}[p]
\epsfxsize=0.95\textwidth
\epsfbox{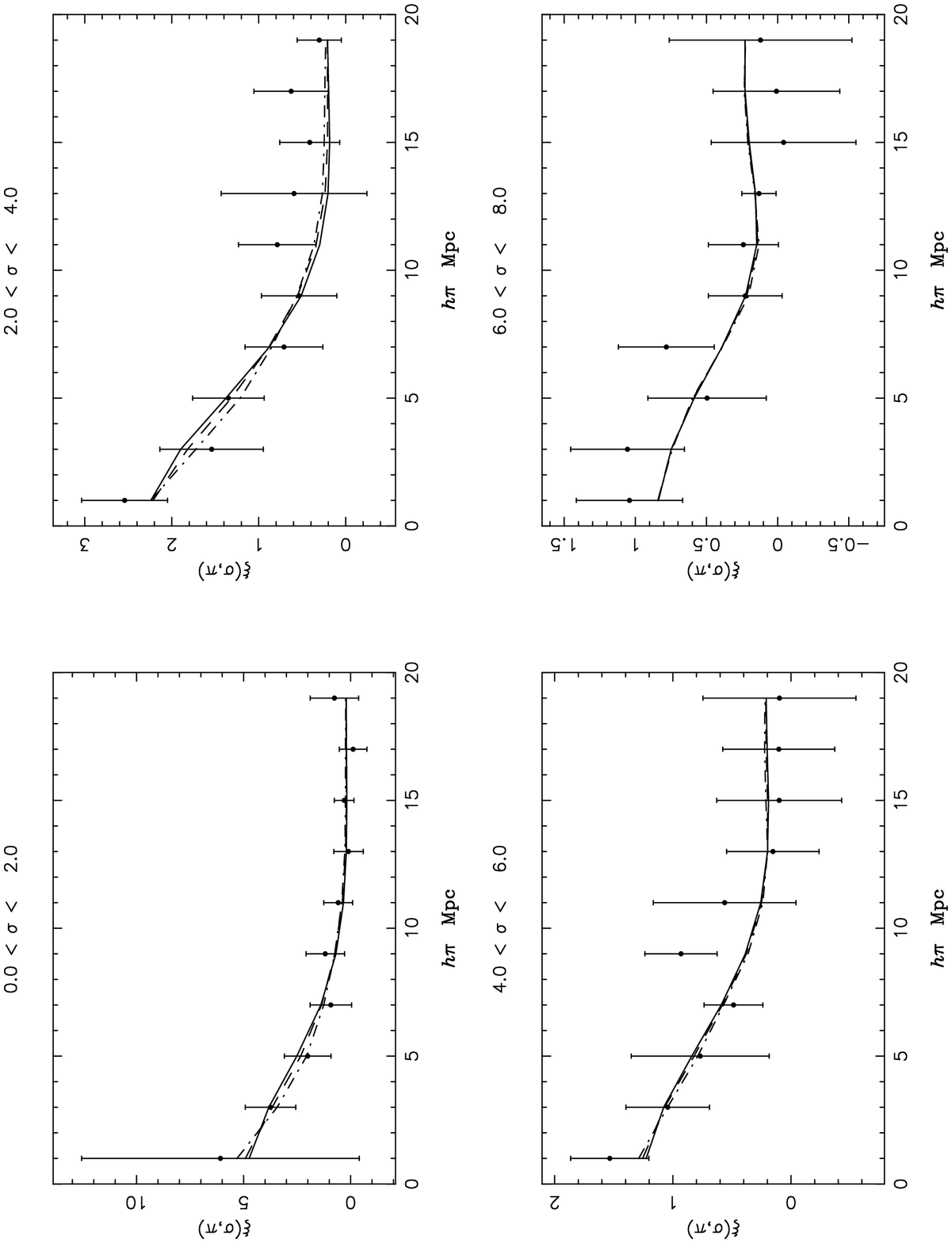}
\caption{a}
\end{figure}

\addtocounter{figure}{-1}
\begin{figure}[p]
\epsfxsize=0.95\textwidth
\epsfbox{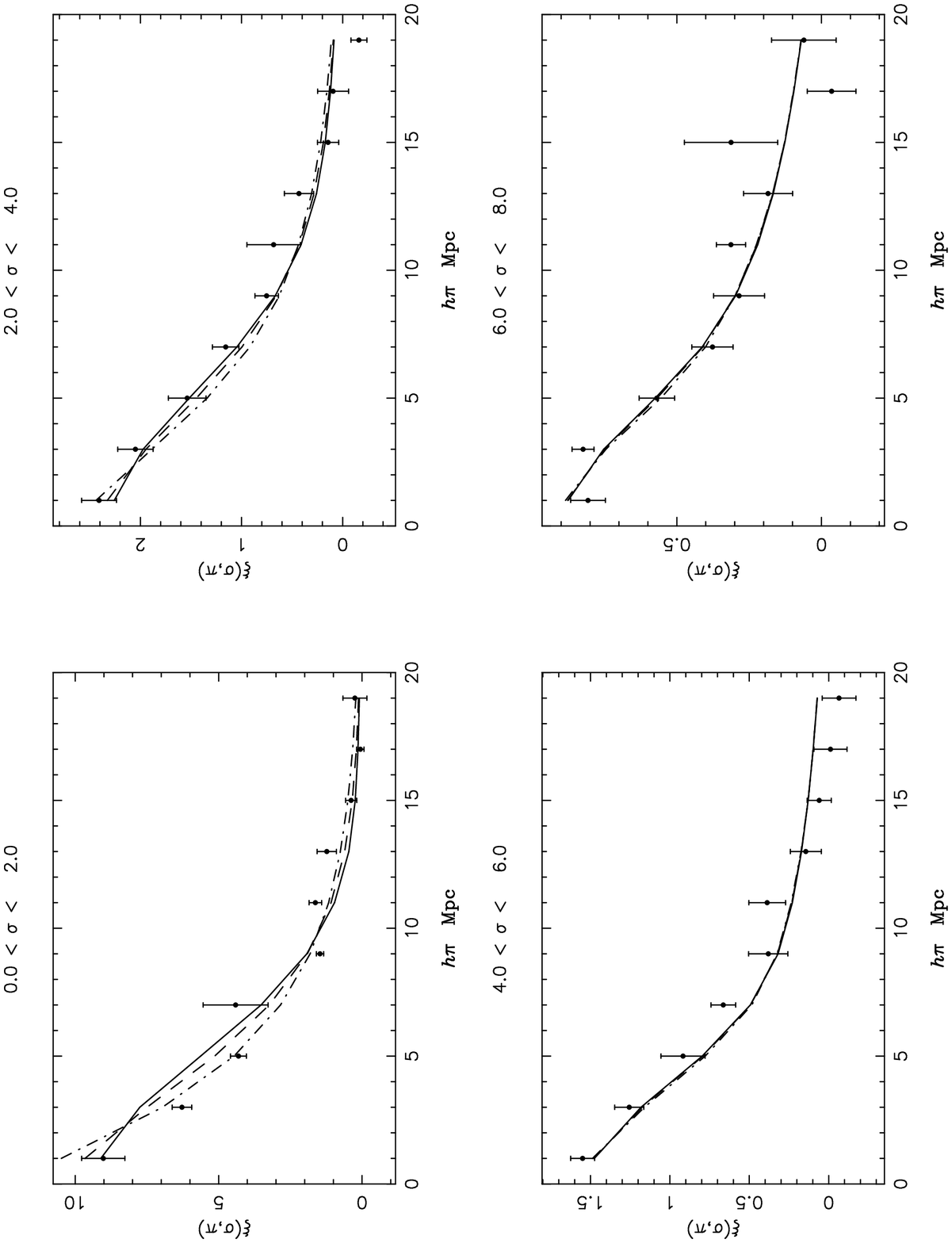}
\caption{b}
\end{figure}

\addtocounter{figure}{-1}
\begin{figure}[p]
\epsfxsize=0.95\textwidth
\epsfbox{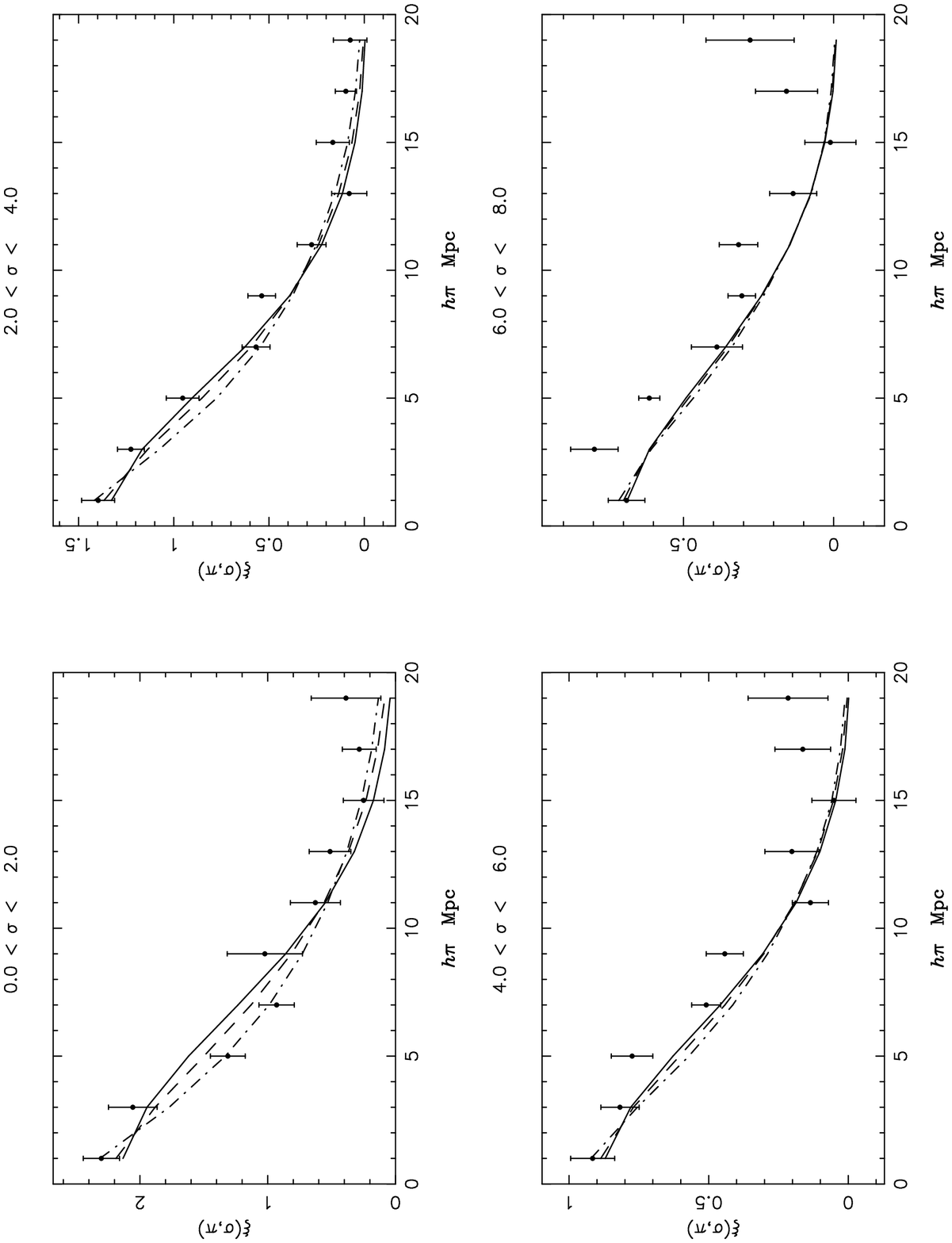}
\caption{c}
\end{figure}

\begin{figure}[p]
\epsfxsize=0.95\textwidth
\epsfbox{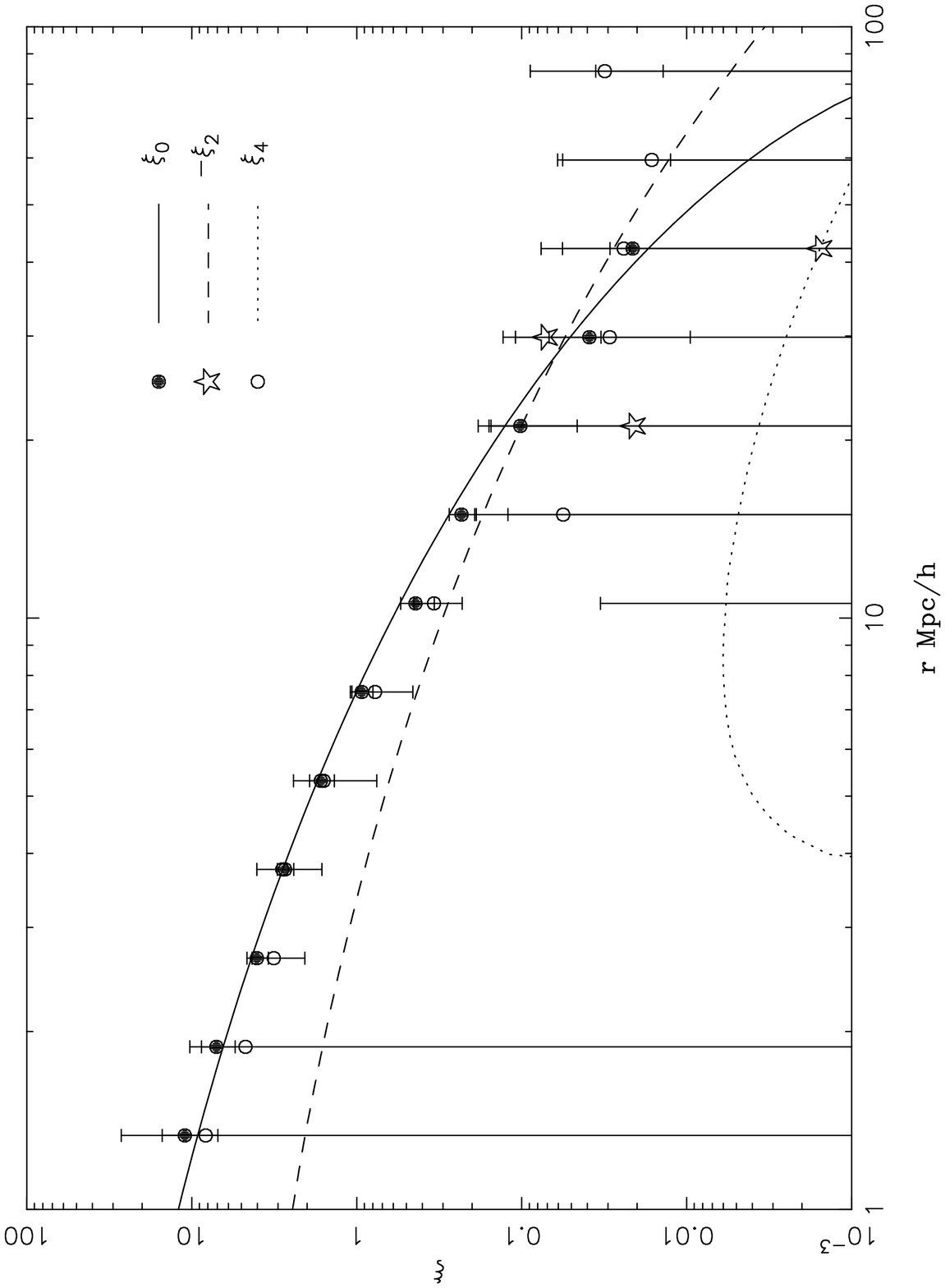}
\caption{a}
\end{figure}

\addtocounter{figure}{-1}
\begin{figure}[p]
\epsfxsize=0.95\textwidth
\epsfbox{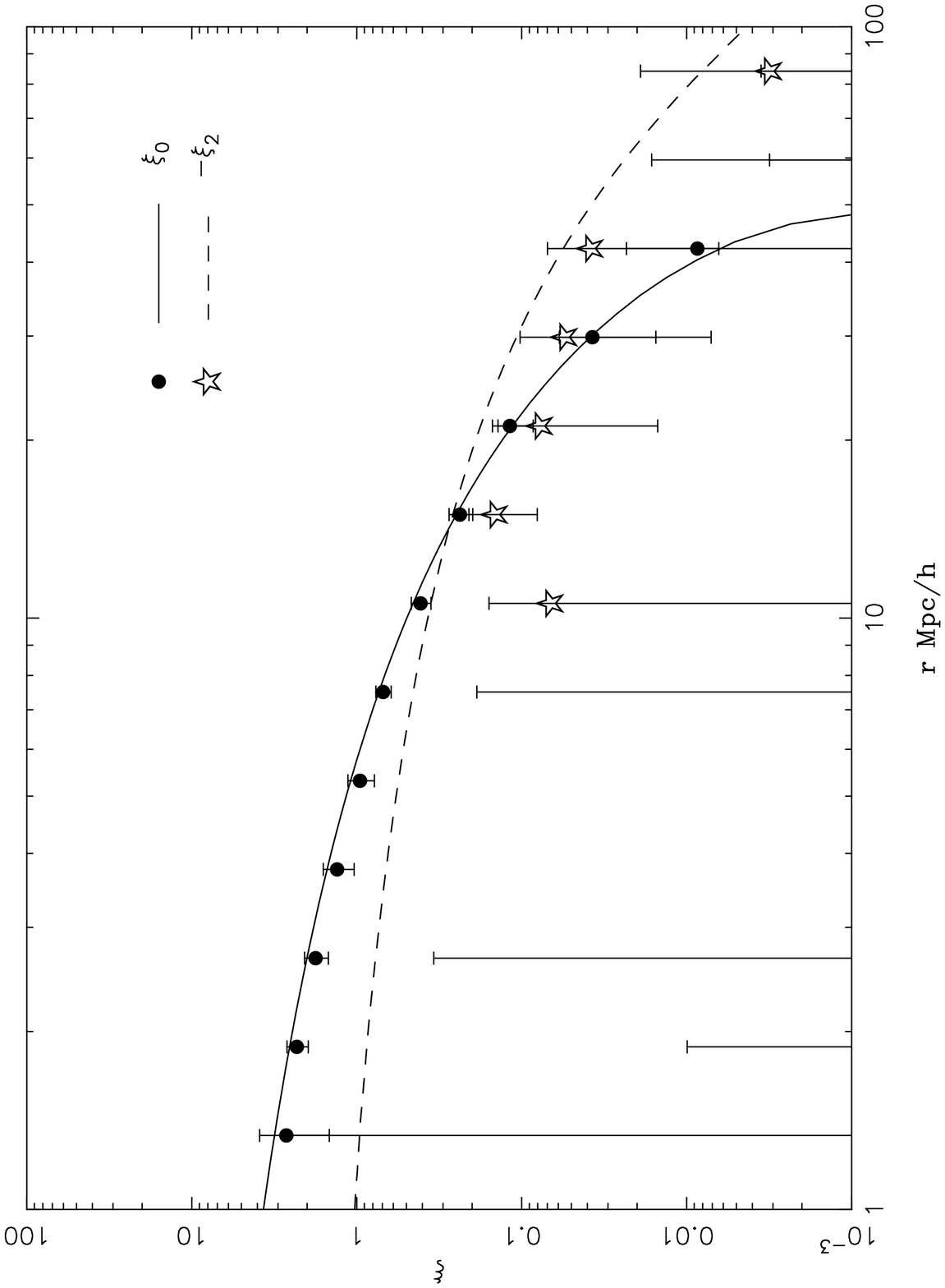}
\caption{b}
\end{figure}

\begin{figure}[p]
\epsfxsize=0.95\textwidth
\epsfbox{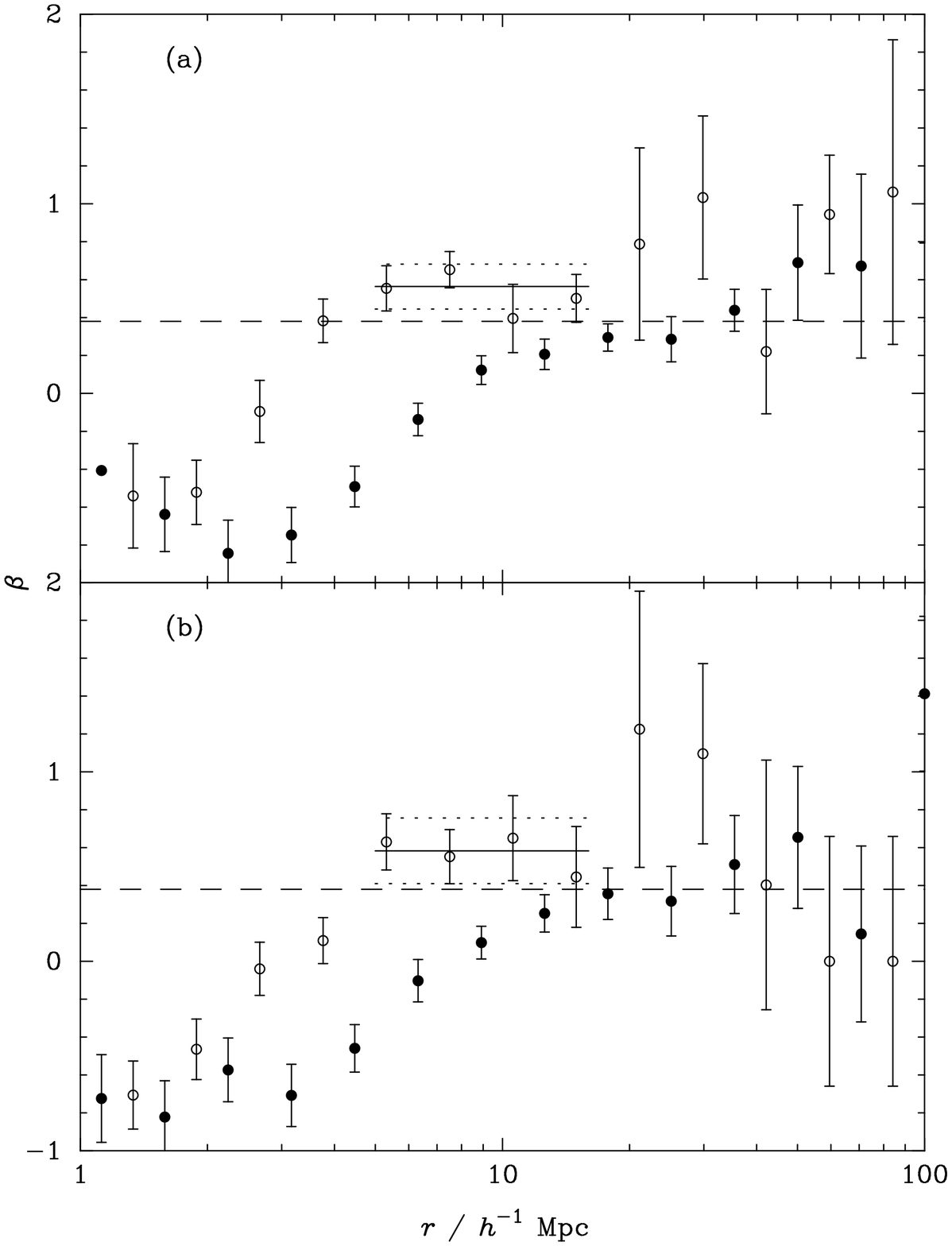}
\caption{\relax}
\end{figure}

\begin{figure}[p]
\epsfxsize=0.95\textwidth
\epsfbox{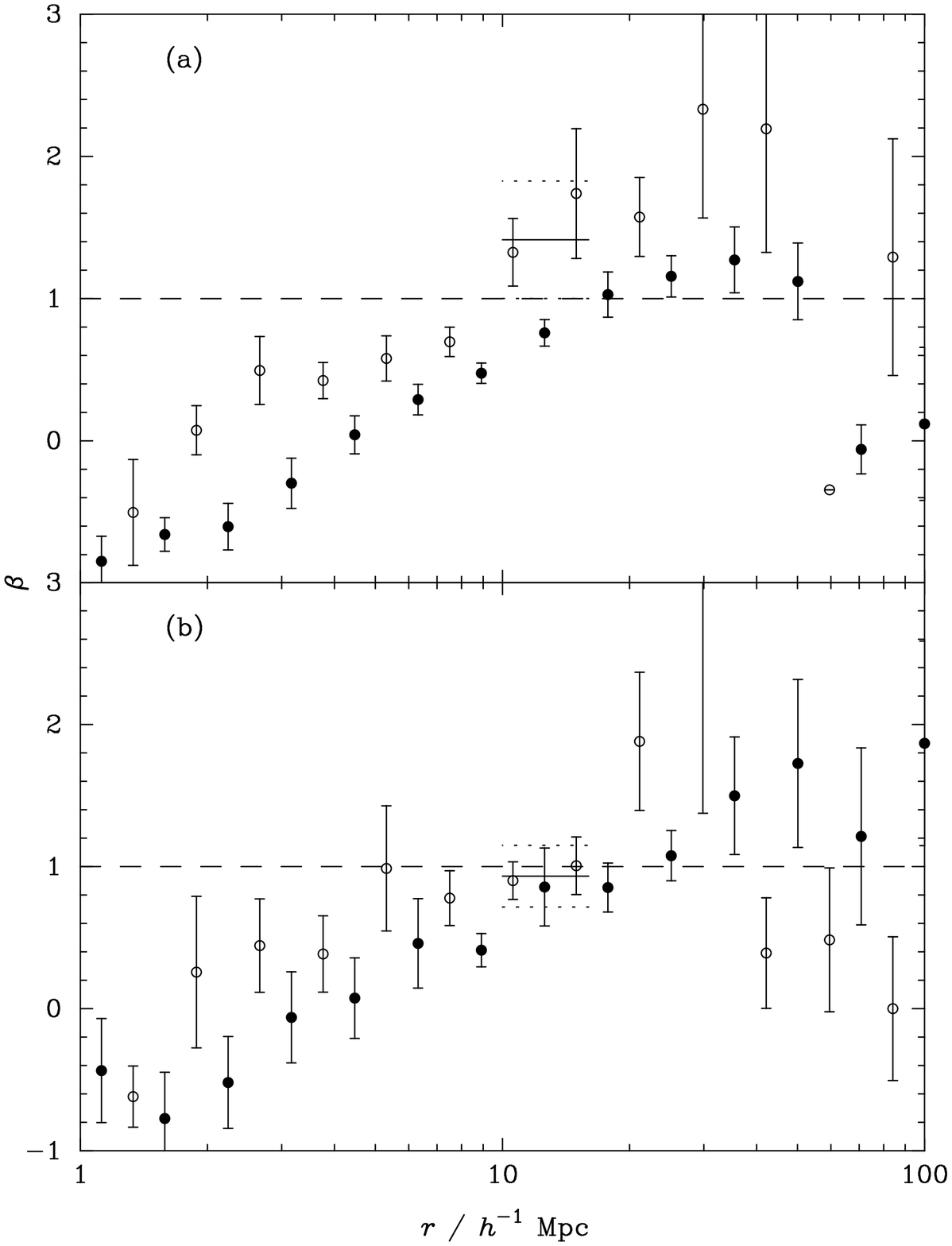}
\caption{\relax}
\end{figure}

\begin{figure}[p]
\epsfxsize=0.95\textwidth
\epsfbox{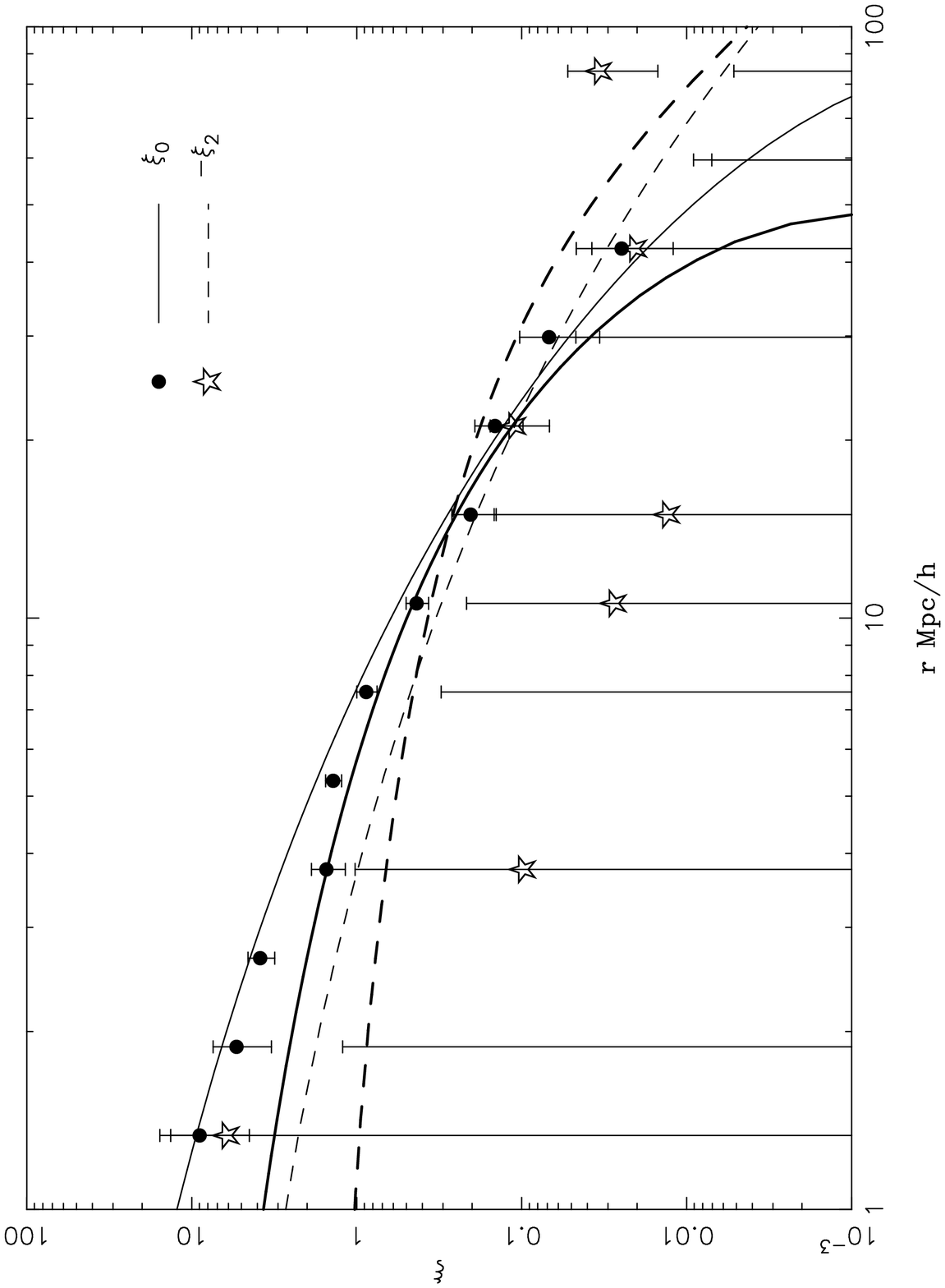}
\caption{\relax}
\end{figure}

\begin{figure}[p]
\epsfxsize=0.95\textwidth
\epsfbox{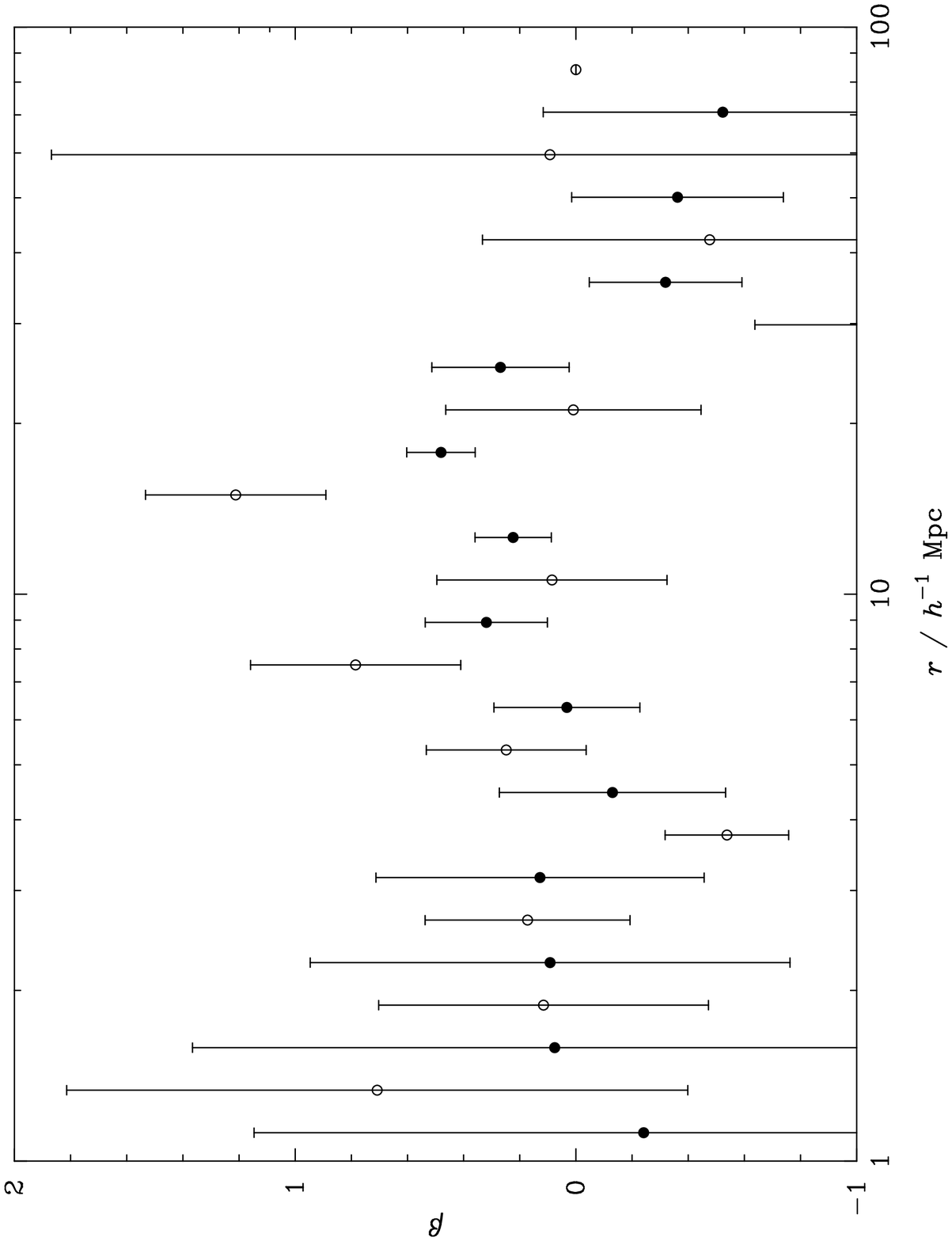}
\caption{\relax}
\end{figure}

\begin{figure}[p]
\epsfxsize=0.95\textwidth
\epsfbox{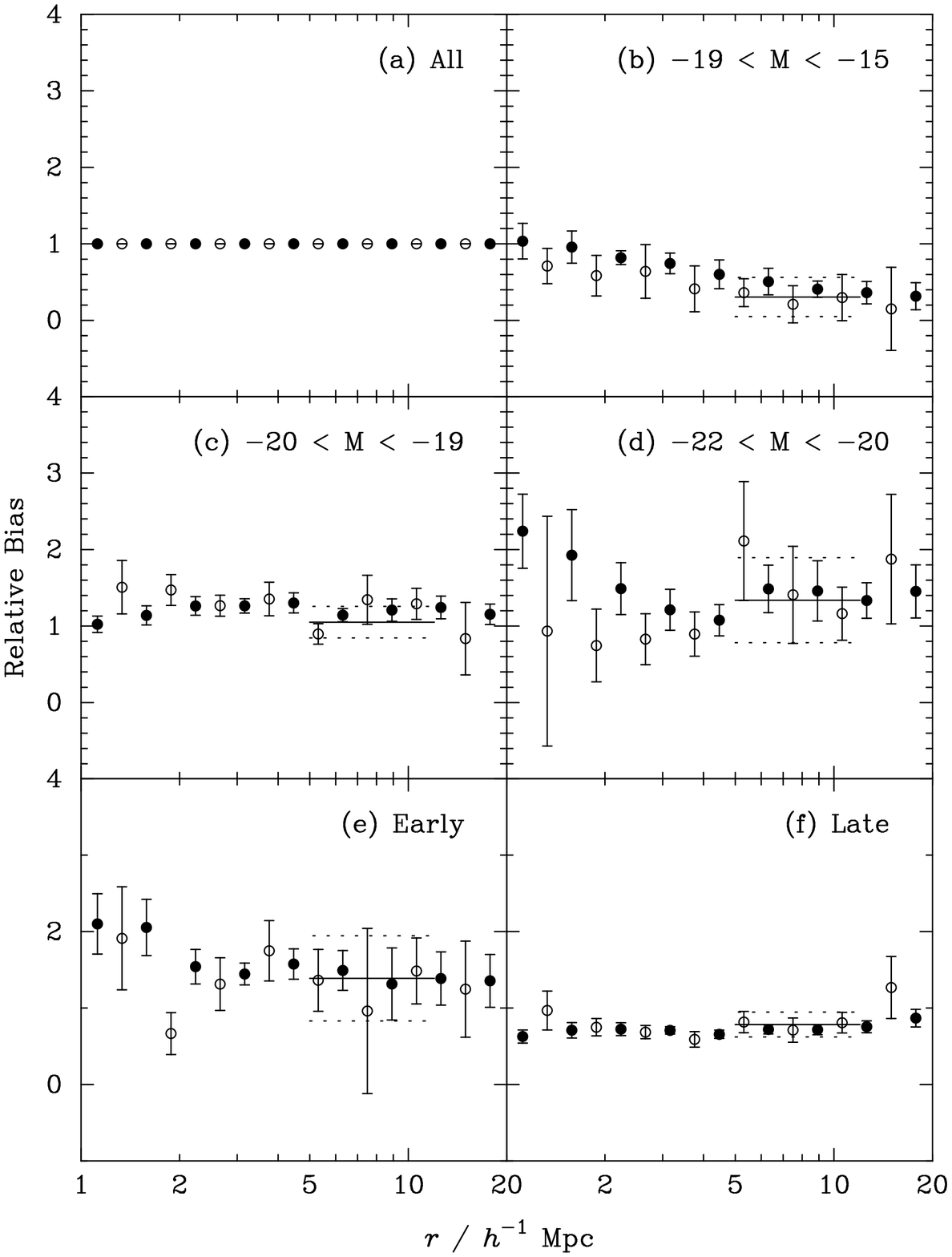}
\caption{\relax}
\end{figure}

\begin{figure}[p]
\epsfxsize=0.95\textwidth
\epsfbox{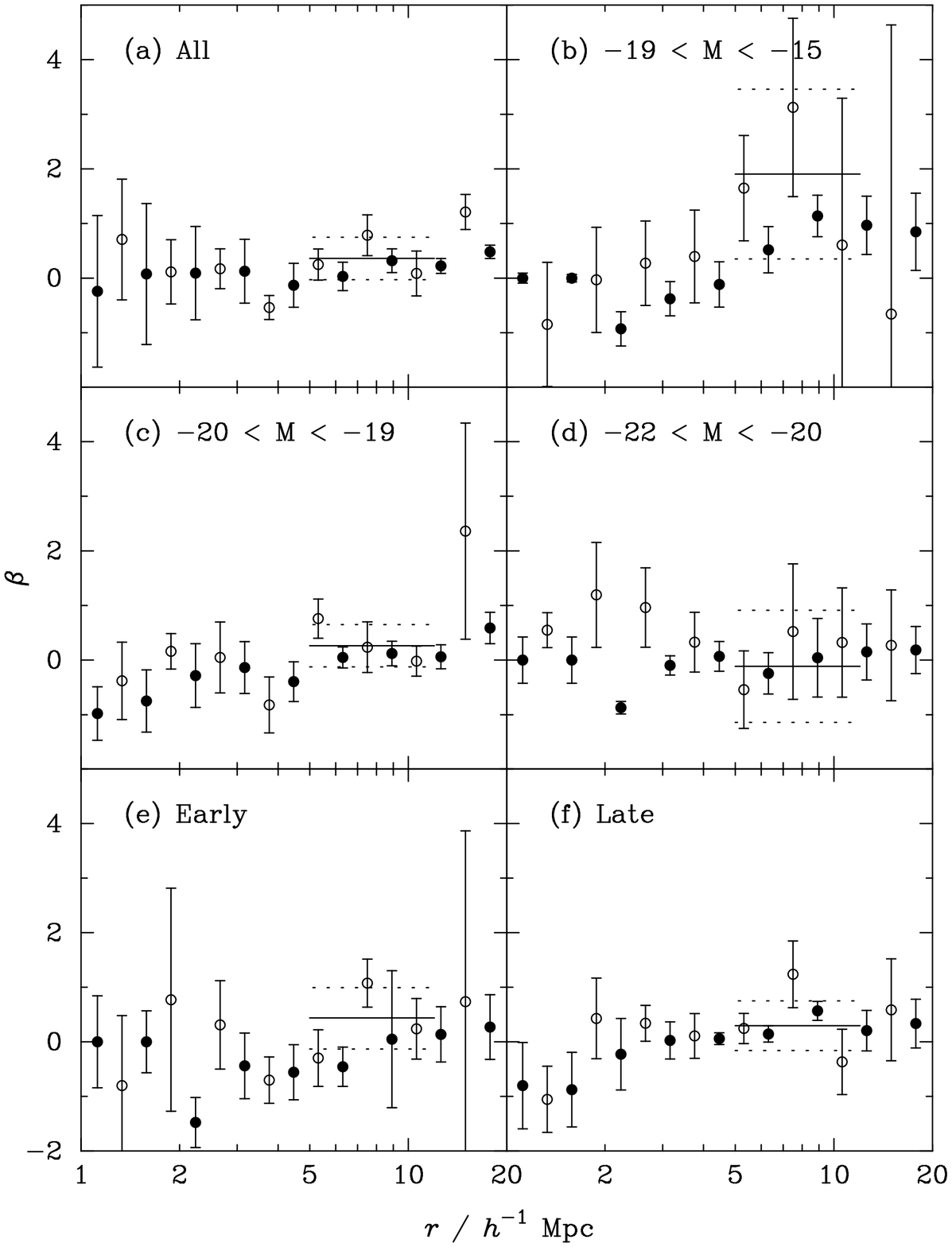}
\caption{\relax}
\end{figure}

\end{document}